\documentclass[aip,reprint,amsmath,amssymb]{revtex4-2}

\usepackage{color}
\usepackage{graphicx}
\usepackage{amsmath}
\usepackage{subcaption}
\usepackage{amssymb}
\usepackage{lineno}
\usepackage{multirow}
\usepackage{natbib}

\newcommand\chg[1]{\textcolor{blue}{#1}} 

\newcommand{\lei}{\lambda_{ei}}

\graphicspath{{figs/}}

\usepackage{placeins}

\begin{document}
	\title{Plasma transport by carrier electrons: delocalization of magnetic field generation}
	
	\author{C. A. Walsh}
	\email{walsh34@llnl.gov}
	\affiliation{Lawrence Livermore National Laboratory, 7000 East Avenue, Livermore, CA 94550}

	\date{\today}
	

	\begin{abstract}
		Magnetic flux generation by Biermann battery occurs more slowly in a kinetic plasma due to the temperature of the current-carrying electrons $T_J$ lagging the bulk plasma temperature $T_e$. VFP simulations of linear temperature perturbations with distinct transverse gradients are used to separate different generation mechanisms: Biermann battery for $\nabla n_e$, composition Sadler for $\nabla Z$, and Kingham-Bell (which has no classical counterpart) for $\nabla \bar{T}_e$. The kinetic stress, originating in the electron distribution's second harmonic, is found to modify these three terms, notably enhancing composition Sadler above its classical generation rate for $Z > 3$. Additional measurements of heat-flux and Nernst advection mean that all $\nabla T_e$ transport terms are characterized simultaneously in a kinetic plasma. Generation terms are more robust to kinetic effects than heat-flux and Nernst; this is in-line with current being carried by slower electrons than those carrying heat. However, they are also relocalized by magnetization more slowly. By using a re-scaled delocalization length ($\lambda_J=0.64\lambda_d$) and delocalization Hall parameter ($\psi_J = 0.30\psi_d$) for the current carriers, suppression of the generation mechanisms follow the suppression function for heat-flux. This work enables self-consistent treatment of energy and magnetic field transport in extended-MHD models.
	\end{abstract}
	\maketitle
	
	\begin{figure}[tb]
		\includegraphics[width=\linewidth]{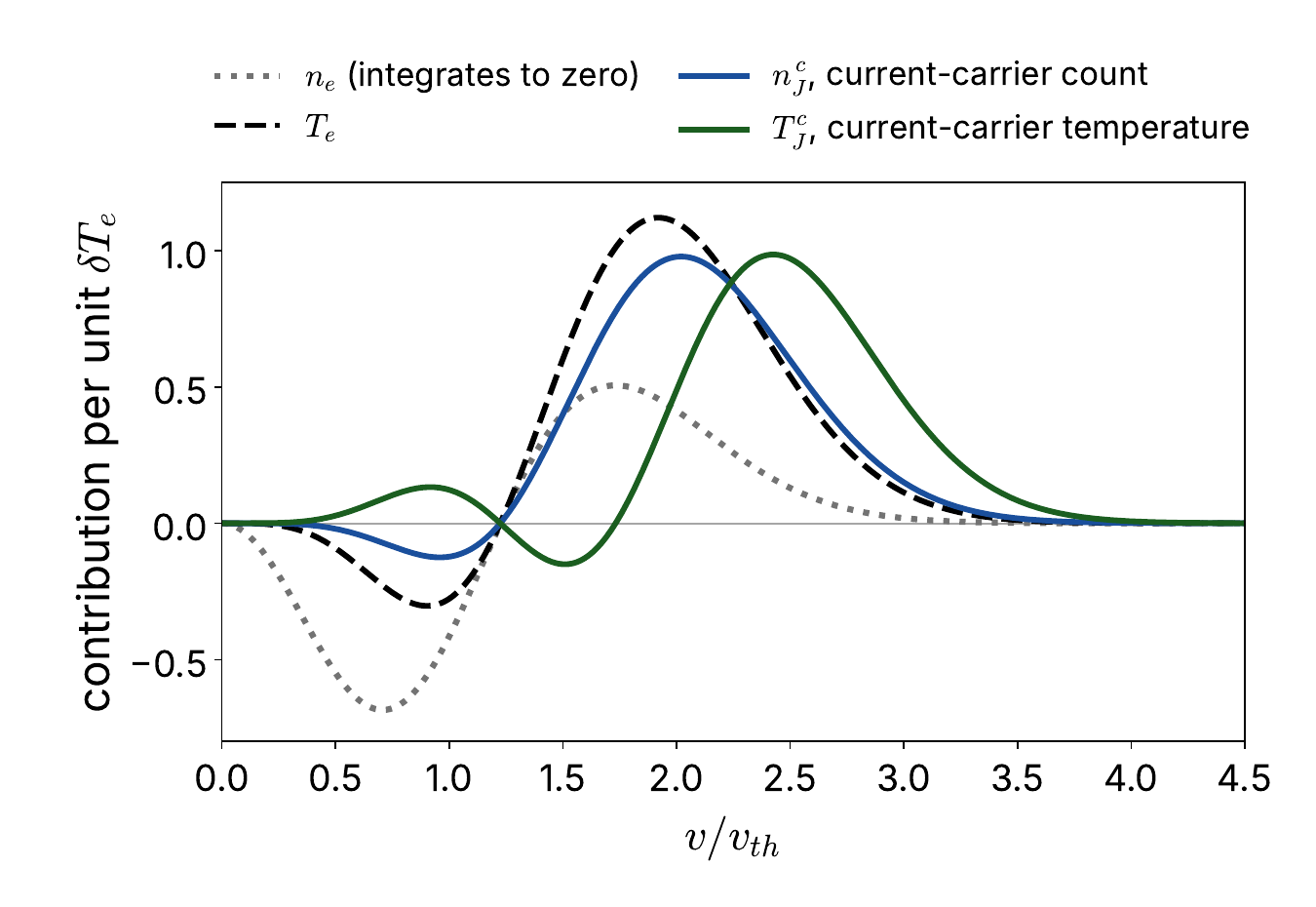}
		\caption{Classical unmagnetized Lorentzian ($Z=\infty$) moments of $f_0$. The temperature and  density seen by the current carrying electrons depends on higher moments than the true density and temperature.  \label{fig:gen_weights} }
	\end{figure}
	
\begin{table*}[tbp]
	\caption{Kinetic transport in a uniform magnetic field, with stress contributions treated separately. Top section: heat-flow\cite{walsh_paper1} and Nernst advection\cite{walsh_paper2} terms. Middle section: B-field generation from the current-carrier terms. Bottom section: kinetic stress, which modifies Nernst, Righi-Leduc and all generation terms. Suppression functions follow $\Gamma_X= 1-S_X[1-P(k_T\lambda_X/(1+\psi_X^{1.6}))]$, where $S_X$ is an amplitude, $\lambda_X$ a length and $\psi_X$ a Hall parameter. With these per-process values, one function applies to all terms: $P(K) = (1 + 27.29K)/(1 + 27.29K + 294.5K^2)$.\cite{walsh_paper1} Stress is first given as the total electric field, before splitting into contributions to each magnetic transport term. Stress and Kingham-Bell are normalized by $\dot{\mathbf B}^c_X = -(T_e/e)\,\nabla\ln X\times\nabla\ln T_e$. In a field the generation stress rows are multiplied by $\mathcal F = {\rm Re}\,(1+i\psi_s)^{-2.64}$. $\hat{\mathbf t}=\nabla T_e/|\nabla T_e|$. $v_{carrier}$ is the peak carrier velocity of each process.}
	\label{tab:gamma}
	\begin{ruledtabular}
		\renewcommand{\arraystretch}{1.75}\small\setlength{\tabcolsep}{2.6pt}
		\begin{tabular}{llcccc}
			Process & Kinetic term & length $\lambda_X$ & Hall parameter $\psi_X$ & amplitude $S_X$ & $v_{carrier}$ \\
			\colrule
			$\parallel$ heat flow
			& $\mathbf q_\parallel = -\Gamma_q\,\kappa_\parallel\nabla_\parallel T_e$
			& $\lambda_d = \dfrac{Z\lambda_{ei}}{\sqrt{Z+4.45}}$ & 0 & 1 & 2.62 \\
			$\perp$ heat flow
			& $\mathbf q_\perp = -\Gamma_q\,\kappa_\perp\nabla_\perp T_e$
			& $\lambda_d$ & $\psi_d = \dfrac{13.4\,Z}{Z+4.45}\,\chi$ & 1 & 2.62 \\
			Righi-Leduc$^\ast$
			& $\mathbf q_\wedge = -\Gamma_q^2\,\kappa_\wedge\,\mathbf b\times\nabla T_e$
			& $\lambda_d$ & $\psi_d$ & 1 & 2.90 \\
			Nernst$^\ddagger$
			& $\mathbf u_N = -\Gamma_N\,\gamma_\perp\nabla_\perp T_e$
			& $1.10\,\lambda_d$ & $\psi_d$ & 1 & 2.69 \\
			Cross-Nernst$^\P$
			& $\mathbf u_{N\wedge} = \Gamma_N^2\,\gamma_\wedge\nabla T_e\times\mathbf b$
			& $1.10\lambda_d$ & $\psi_d$ & 1 & 2.96 \\
			\colrule
			\multicolumn{6}{l}{\emph{Generation, current-carrier part}, \ $\sigma(Z) = \frac{Z+1.54}{Z+3.69}$} \\
			Biermann ($\nabla n_e$)
			& $\dot{\mathbf B}_n = \Gamma_B (\nabla T_e \times \nabla n_e)/(e n_e)$
			& $\lambda_J = 0.64\,\lambda_d$ & $0.30\,\psi_d$ & $0.75\,\sigma(Z)$ & 2.43 \\
			Comp. Sadler ($\nabla Z$)
			& $\dot{\mathbf B}_Z = \Gamma_Z\,\beta_\parallel'(Z)\,\nabla Z\times\nabla T_e /e$
			& $\lambda_J$ & $0.30\,\psi_d$ & $\sigma(Z)$ & 2.02 \\
			Kingham-Bell ($\nabla \bar{T}$)
			& $\dot{\mathbf B}_{\bar{T}} = (\Gamma_{\rm KB} - 1)\,\dot{\mathbf B}^c_{\bar T}$
			& $\lambda_J$ & $0.30\,\psi_d$ & $\sigma(Z)$ & both \\
			\colrule
			\multicolumn{6}{l}{\emph{Stress}, \ $\psi_s = \dfrac{20.8\,Z\chi/(Z+6.6)}{1+k_T\lambda_d/0.105}$, \ $s=S_X (1-P(k_T\lambda_X))$} \\
			Total stress field
			& $e(E_{\Pi\parallel} + iE_{\Pi\wedge}) = \,T_e s(1-i\psi_s)^{-1.7}\partial_\parallel\ln T_e$
			& $0.645\,\lambda_d$ & $\psi_s$ & $\dfrac{2.95}{Z+6.3}$ & 3.04 \\
			\multicolumn{6}{l}{\quad\emph{Stress contributions to $\partial\mathbf B/\partial t$:}} \\
			\quad Nernst
			& $\Delta\Gamma_N = -\,{\rm Im}\big[s(1-i\psi_s)^{-1.7}\big]/\beta_\wedge^c$
			& --- & $\psi_s$ & --- & 3.3 \\
			\quad Biermann ($\nabla n_e$)
			& $\dot{\mathbf B}_{\Pi,n} = \dot{\mathbf B}^c_{n}\left[\tfrac32\,\partial s/\partial\ln n_e + 0.16\,s\right]$
			& --- & $\psi_s$ & --- & --- \\
			\quad Comp.\ Sadler ($\nabla Z$)$^\#$
			& $\dot{\mathbf B}_{\Pi,Z} = \dot{\mathbf B}^c_{Z}\left[\tfrac32\,\partial s/\partial\ln Z + 0.47\,w\,h\,s\right]$
			& --- & $\psi_s$ & --- & --- \\
			\quad Kingham-Bell ($\nabla\bar T$)
			& $\begin{aligned}\dot{\mathbf B}_{\Pi,\bar T} = -\dot{\mathbf B}^c_{\bar T}\big[&0.38\,s + 0.69\,\partial s/\partial\ln n_e\\[-4pt] &+ 0.26\,(1-P(0.1\,k_T\lambda_d))\big]\end{aligned}$
			& --- & --- & --- & --- \\
		\end{tabular}
	\end{ruledtabular}
	\raggedright\footnotesize
	$^\ast$ Uniform-field total\cite{walsh_paper1}, including the stress contribution. In a non-uniform field Righi-Leduc also depends on  field wavelength\cite{walsh_paper2}.
	$^\ddagger$ Purely advective part; the total is $\Gamma_N + \Delta\Gamma_N$ (stress row). The previous paper\cite{walsh_paper2} additionally separates the field-induced change of $f_0$ ($S_{f0}$), which gives Nernst relocalizing with $0.78\,\psi_d$. 
	$^\P$ Cross-Nernst, written here as $\propto\Gamma_N^2$, can also be written as $\Gamma_q^2$.\cite{walsh_paper2}.
	$^\#$ Cross shear, fitted: $w = Z/(Z+12.2)$, $h = 1-(1+2.2\,w)\,(1-P(0.57\,k_T\lambda_d))$.
\end{table*}

	Plasmas with sharp temperature gradients have suppressed electron heat-flow due to kinetic effects\cite{malone1975}. The electron distribution becomes non-Maxwellian, with depletion of the less-collisional fast electrons reducing the effective thermal conductivity\cite{bell1981,luciani1983,epperlein1991,walsh_paper1}. But heat-flow is not the only transport process that depends on the electron distribution. 
	
	Theoretical work has shown that electron kinetics impacts transport of magnetic fields down temperature gradients\cite{brodrickTestingNonlocalModels2017,brodrickIncorporatingKineticEffects2018,sherlockSuppressionBiermannBattery2020,walshKineticCorrectionsHeatflow2024}, as well as the generation of fields at temperature gradients\cite{sherlockSuppressionBiermannBattery2020,daviesNonlocalSuppressionBiermann2023}. Experiments are now capable of benchmarking these kinetic models\cite{campbell2021measuring,griff-mcmahon2024,arran2025}. However, there is no theoretical framework that simultaneously handles all transport processes. Studies have been limited to documenting individual terms rather than unifying all kinetic plasma transport. Often they have used unconverged electron distributions\cite{daviesNonlocalSuppressionBiermann2023,liu2017} or lack full treatment of electron collisions\cite{daviesNonlocalSuppressionBiermann2023,kingham2002}. 
	
	The importance of a unified kinetic transport model is clear when considering the interweaving contribution of each process on a strong temperature gradient $\nabla T_e$. Heat-flux lessens $\nabla T_e$; magnetization sharpens $\nabla T_e$ by reducing heat-flux; Nernst advection reduces magnetization, thereby smoothing $\nabla T_e$\cite{hainesHeatFluxEffects1986}; Field generation magnetizes the gradient, sharpening $\nabla T_e$ and damping Nernst. Accurately assessing the relative strength of each process is critical for a transport model\cite{ridgersMagneticCavitationReemergence2008}. 

	This is the final paper in a 4-part series studying plasma transport of magnetic fields and electron energy in the linear regime. The question tying the series together is: 'which electron speeds carry the transport?' The first paper used this concept to explain the behavior of the classical transport coefficients across ionization and magnetization\cite{walsh_paper0}. The second focused on  kinetic heat-flux\cite{walsh_paper1}, measuring the delocalization length $\lambda_d$ and the delocalization Hall parameter $\psi_d$. The third paper studied Nernst advection of magnetic fields down temperature gradients\cite{walsh_paper2}, showing how kinetic stress\cite{liu2015,liu2017} (which has been overlooked in extended-MHD codes) results in Nernst being more suppressed than heat-flow.
	
	This paper completes the series by applying the same methodology to the generation of magnetic fields. For the first time, all energy and magnetic transport is simultaneously measured and characterized from the linear response of a kinetic plasma. The simple truth that enables this work is:
	\begin{quote}
		\emph{All transport comes from the same electron distribution, but each process reads it differently.}
	\end{quote}

	The paper is organized as follows. Section \ref{sec:theory} provides the theoretical foundation, starting with the classical induction equation (section \ref{sec:theory_classical}) before reviewing the key kinetic metrics (delocalization length $\lambda_d$ and delocalization Hall parameter $\psi_d$) in section \ref{sec:theory_kinetic}. Section \ref{sec:theory_currentcarriers} introduces the  current-carrier temperature and density ($T_J$, $n_J$); the difference between these and their classical quantities (e.g. $T_e$) describe kinetic suppression of Biermann battery generation (section \ref{sec:theory_bier}), composition Sadler generation\cite{sadlerMagneticFieldGeneration2020} (section \ref{sec:theory_sadler}), and the emergence of Kingham-Bell generation\cite{kingham2002} (section \ref{sec:theory_kingham}). Section \ref{sec:theory_stress} introduces an additional Ohm's law term that requires the second harmonic ($\underline{\underline{f}}_2$) of the electron distribution to be retained: field generation by kinetic stress. 
		
	The VFP simulations used to measure the suppression functions are discussed in section \ref{sec:sims}. Section \ref{sec:gen_current_carriers} presents the results of the current-carrier portion of generation suppression, while section \ref{sec:gen_stress} incorporates the stress contribution. Combined results from this paper series are given in section \ref{sec:discussion}, through the kinetically-corrected induction and heat-flow equations; the relevant information for those two equations is given in table \ref{tab:gamma}.
	
	Appendix \ref{sec:harmonic_trunc} quantifies the error in each transport process through truncating the electron distribution; this is evidence that this paper uses converged harmonics. Appendix \ref{sec:davies} compares the work of this paper to previous work characterizing suppression of Biermann battery\cite{sherlockSuppressionBiermannBattery2020,daviesNonlocalSuppressionBiermann2023}. Under the same assumptions, the data from Davies\cite{daviesNonlocalSuppressionBiermann2023} is reproduced, validating the computational machinery here; the importance of including converged harmonics and a full collision operator is then demonstrated. 
	
	\section{Theory \label{sec:theory}}
	
	\subsection{Classical induction equation \label{sec:theory_classical}}
	In a classical plasma, the magnetic field is transported as follows\cite{braginskiiTransportProcessesPlasma1965,epperleinPlasmaTransportCoefficients1986,walshExtendedmagnetohydrodynamicsDensePlasmas2020,sadlerSymmetricSetTransport2021}:
	\begin{align}
		\begin{split}
			\frac{\partial \mathbf{B}}{\partial t} = &\nabla \times \Bigg( \frac{\nabla P_e}{e n_e} + \frac{\beta_{\parallel} \nabla T_e}{e}\Bigg) \\ 
			&+ \nabla \times \Big((\mathbf{v}_{plasma}+\mathbf{v}_{N}) \times \mathbf{B} \Big)
			\label{eq:mag_trans_new}
		\end{split}
	\end{align}
	\begin{equation}
		\label{eq:mag_trans_vnernst}	\mathbf{v}_{N} = - \gamma_{\bot} \nabla T_e - \gamma_{\wedge}\hat{\mathbf{b}} \times \nabla T_e 
	\end{equation}

	The first line of equation \ref{eq:mag_trans_new} is generation of magnetic fields; the second is advection with both the bulk flow ($v_{plasma}$) and the Nernst velocity $\mathbf{v}_N$ down temperature gradients.  
	
	The first field-generating term in equation \ref{eq:mag_trans_new} is the Biermann battery. Using the ideal gas approximation ($P_e = n_e T_e$) shows the effect more obviously: $\nabla T_e \times \nabla n_e/en_e$; field generation at misaligned temperature and electron density gradients. 
	
	The second field-generating term can be understood by using the identity $\nabla \times \nabla T_e = 0$. Classically, $\beta_\parallel$ only depends on ionization, leaving $(\partial \beta_\parallel / \partial Z) \nabla Z \times \nabla T_e$; field generation at misaligned temperature and ionization gradients. This term is named composition Sadler generation \cite{sadlerMagneticFieldGeneration2020}.
	
	Magnetization is measured by the Hall parameter $\chi=\omega_{ce}\tau_B$, corresponding to the magnetization of the electrons traveling at $1.10v_{th}$, due to  Braginskii's\cite{braginskiiTransportProcessesPlasma1965} definition of the collision time $\tau_B=3\sqrt{\pi}\tau_{ei}/4$. The thermal electron velocity is defined as $v_{th} = \sqrt{2T_e/m_e}$. The electron-ion mean free path is $\lambda_{ei}=v_{th}\tau_{ei}$:
	\begin{equation}
		\lambda_{ei} \;=\;  \frac{16 \pi \epsilon_0^2 T_e^2}{n_e Z e^4 \ln\Lambda} 
	\label{eq:lambdaei}
	\end{equation}
	
	Classical transport coefficients written with a super-script '$c$' are dimensionless (e.g. $\gamma_{\bot}^c$) and only depend on ionization ($Z$) and the Hall parameter ($\chi$). 

	Current-driven transport has been neglected for this work, although resistive diffusion should be considered in many laser-driven plasmas\cite{hohenbergerInertialConfinementFusion2012,manuelStudyHydrodynamicInstabilities2021,walsh2024}. 
	
	Ohm's law also contains the divergence of the electron stress, $-\nabla\cdot\underline{\underline{\Pi}}_e/(en_e)$. Classically the stress is driven by electron flow\cite{braginskiiTransportProcessesPlasma1965,catto2005,mikhailovskii1984} so it vanishes with the ions at rest and no current, and it is omitted from equation~\ref{eq:mag_trans_new}. Section~\ref{sec:theory_stress} shows that a kinetic plasma carries a stress driven by heat flow rather than electron flow.
	
	\subsection{From classical to kinetic transport \label{sec:theory_kinetic}}
	
	The electron distribution is expanded in spherical harmonics, $f_0(v)$, $\mathbf{f}_1(v)$,
	$\underline{\underline{f}}_2(v)$, \ldots, where $v$ is the electron speed\cite{bell1981,tzoufrasVlasovFokkerPlanck2011,shkarofskyip1966}.
	$f_0$ holds the bulk properties (density, temperature), $\mathbf{f}_1$ the fluxes, and
	$\underline{\underline{f}}_2$ the anisotropic temperatures ($T_{xx}\ne T_{yy}$) and shear ($T_{xy}\ne0$).
	Classical transport keeps $f_0$ Maxwellian and truncates at $\mathbf{f}_1$.
	
	Kinetic effects begin when the carrier electrons travel further between collisions than the gradient length. For heat-flow the relevant distance is the delocalization length\cite{walsh_paper1}
	\begin{equation}
		\lambda_d = \lambda_{ei} \frac{Z}{\sqrt{Z+4.45}} 
		\label{eq:lam_d}
	\end{equation}
	The $Z$-dependence comes from electron-electron (e-e) collisions, which slow the carrier population at low $Z$\cite{walsh_paper0}. A magnetic field relocalizes the
	delocalized electrons through the delocalization Hall parameter\cite{walsh_paper1}:
	\begin{equation}
		\psi_d = 13.4\,\chi \frac{Z}{Z+4.45} \label{eq:psid}
	\end{equation}
	which lowers the delocalization length through: 
	\begin{equation}
		\lambda_{d\perp} = \frac{\lambda_d}{1+\psi_d^{1.6}} \label{eq:lam_dperp}
	\end{equation}
	For a temperature perturbation of wavenumber $k_T$ the heat flow is then reduced to $\Gamma_q = P(k_T\lambda_{d\perp})$ of its classical value at every $Z$:\cite{walsh_paper1}
	\begin{equation}
		P(K) = \frac{1+27.29K}{1+27.29K+294.5K^2}
		\label{eq:P}
	\end{equation}
	
	Every process in this paper is written the same way: its classical value multiplied by a suppression function
	$\Gamma_X$, with $\Gamma_X=1$ the classical limit. The functions take the form
	$\Gamma_X = 1 - S_X\left[1-P\left(k_T\lambda_X/(1+\psi_X^{1.6})\right)\right]$, where the length $\lambda_X$ and
	Hall parameter $\psi_X$ say which electrons carry the process and $S_X$ how strongly the process depends on them.
	
	\subsection{Current carriers in the induction equation\label{sec:theory_currentcarriers}}
	
	A steady-state ($\partial \mathbf{f}_1/\partial t = 0$) plasma with ions at rest ($v_{plasma}=0$) and no current ($\mathbf{J}=0$) is assumed. The zero current assumption may seem counter-intuitive, as field generation inherently creates current. However, the transport is then assessed relative to the bulk electron velocity ($-\mathbf{J}/(en_e)$), which is the well-known Hall term.
	
	The electric field is a direct consequence of the zero current condition; it is the electric field required to have no net electron drift. The magnetic transport is then found through $\partial \mathbf{B}/\partial t = - \nabla \times \mathbf{E}$. While the net current in the plasma is taken as zero, the current for a given electron speed is non-zero; the sum over the population equals zero. 
	
	The electric field, assuming harmonic truncation at $\mathbf{f}_1$, is:
	\begin{equation}
		\mathbf{E} = - \frac{\nabla(n_{J} T_{J})}{en_{J}} - \mathbf{v}_N \times \mathbf{B} \label{eq:E_kinetic_origin}
	\end{equation}
	which is true for arbitrary $f_0$.\cite{hainesHeatFluxEffects1986,shkarofsky1979}.

	$n_J$ and $T_J$ are the density and temperature seen by the current-carriers (moments of $f_0$ weighted by the current-carrying electrons). In the Lorentz limit ($Z=\infty$) the terms are simple moments of the electron distribution\cite{shkarofsky1979}:
	\begin{equation}
		n_{JL} \equiv n_e \langle v^3 \rangle \label{eq:E_kinetic_origin_n}
	\end{equation}
	\begin{equation}
		T_{JL} \equiv \frac{m_e}{6}\frac{\langle v^5 \rangle}{\langle v^3 \rangle}  \label{eq:E_kinetic_origiT}
	\end{equation}
	Where the moments of $f_0$ are defined as $\langle v^m\rangle\equiv(4\pi/n_e)\int_0^\infty f_0\,v^{m+2}\,\mathrm dv$. The subscript '$JL$' denotes that $T_{JL}$ is the Lorentzian temperature seen by the current carriers. 
	
	The form of equation \ref{eq:E_kinetic_origin} has been chosen to emphasize the kinetic origin of equation \ref{eq:mag_trans_new}. Once again, the first term is generation of magnetic fields and the second is advection with Nernst velocity $\mathbf{v}_N$. The moment definition of $\mathbf{v}_N$ can be found in the previous paper\cite{walsh_paper2}, which studied the kinetic modification of Nernst. This paper focuses on field generation:
	
	\begin{equation}
		\Bigg[\frac{\partial \mathbf{B}}{\partial t}\Bigg]_{gen} = \nabla \times \frac{\nabla (n_J T_J)}{e n_J} \label{eq:Bgen_kin}
	\end{equation}
	
	The key to understanding kinetic modification of field generation is the difference between the true plasma conditions ($T_e$, $n_e$) and the conditions of the current-carrying electrons ($T_J$, $n_J$). A Maxwellian plasma at any ionization gives $T_J = T_e$. The density of the current-carrying electrons, on the other hand, \emph{is} dependent on ionization; for a $Z=\infty$ plasma:
	\begin{equation}
		n_{JL}^c \propto n_e T_e^{3/2} \label{eq:nJLmax}
	\end{equation} 
	
	For arbitrary $Z$ the classical current carrier density ($n_J^c$) is:
	\begin{equation}
		n_{J}^c \propto n_e T_e^{\beta_{\parallel}} \label{eq:nJmax}
	\end{equation}
	
	Note that, for ease, this paper describes $T_J$ and $n_J$ as the temperature and density of the current-carrying electrons, while it is more accurately described as the temperature and density \emph{inferred} by their part of the distribution.
	
	Sections \ref{sec:theory_bier} and \ref{sec:theory_sadler} show how the Biermann battery and composition Sadler terms (of classical equation \ref{eq:mag_trans_new}) emerge from kinetic equation \ref{eq:Bgen_kin}. Section \ref{sec:theory_kingham} then describes a term originating in equation \ref{eq:Bgen_kin} that has no classical counterpart; this term was first observed by Kingham \& Bell\cite{kingham2002}, but remains uncharacterized across $Z$ and $k_T$. Section \ref{sec:theory_stress} then presents a term omitted from equation \ref{eq:Bgen_kin}: the kinetic heat-flux stress. 
	
	\subsubsection{Kinetic origin of Biermann battery generation ($\nabla n_e$) \label{sec:theory_bier}}

	For a Maxwellian plasma in the Lorentz limit ($Z=\infty$) $T_{JL} = T_e$:
	\begin{equation}
		\mathbf{E}_{gen}= - \frac{\nabla P_e}{en_e} - \frac{3}{2} \frac{\nabla T_e}{e}, \qquad P_e = n_eT_e, \label{eq:Egen_bier}
	\end{equation}
	

	Once the curl is taken, the $\nabla T_e$ term drops out, due to $\nabla \times \nabla T_e = 0$. What remains is the classical Biermann battery $(\nabla n_e \times \nabla T_e)/en_e$. 
	
	The kinetic response of Biermann battery is simply: 'how well does $T_{J}$ track $T_e$'? $T_e = (m_e/3) \langle v^2 \rangle$, while the current carrying electrons see (for a Lorentz plasma) $T_{JL}=(m_e/6) \langle v^5 \rangle / \langle v^3 \rangle$; $T_{JL}$ depends on higher velocity electrons, even though it equals $T_e$ classically.  As a plasma is heated, $T_e$ rises, but $T_J$ lags due to hot (less collisional) electrons escaping the region. 
	
	Kinetic suppression of Biermann battery is therefore the lag of the current-carrier temperature:
	\begin{equation}
		\Gamma_B = \Gamma_{TJ} = \frac{\delta T_J}{\delta T_e}, \label{eq:Gam_B}
	\end{equation}
	
	e-e collisions for finite $Z$ plasmas lower the velocity of the current-carrying electrons\cite{walsh_paper0}, allowing $T_J$ to track the true temperature more closely. Therefore, Biermann battery is less suppressed at low $Z$. This is where the factor $\sigma(Z)=(Z+1.54)/(Z+3.69)$ in equation \ref{eq:gamma_B_fit} originates. 
	
	\subsubsection{Kinetic origin of composition Sadler generation ($\nabla Z$) \label{sec:theory_sadler}}
	
	In the previous section, the term $\nabla \times (3/2) \nabla T_e$ was dropped from equation \ref{eq:Egen_bier} due to $\nabla \times \nabla T_e = 0$.  More generally, the $3/2$ factor is $\beta_{\parallel}(Z=\infty)$. Therefore, in a plasma with varying ionization, the term is not dropped.
		
	The Maxwellian relation $n_J^c \propto n_e T_e^{\beta_{\parallel}}$ indicates that a hotter plasma has more electrons at the current-carrying speed. $\beta_{\parallel}$ is the rate at which the number of current-carriers grows as the plasma is heated\cite{walsh_paper0} ($\beta_\parallel = \partial\ln n_J/\partial\ln T_e$), with $\beta_{\parallel}(Z=\infty)=3/2$ and $\beta_{\parallel}(Z=1)=0.71$.\cite{epperleinPlasmaTransportCoefficients1986,sadlerSymmetricSetTransport2021} Equivalently, it is the mean energy of the current carriers above the enthalpy, $\beta_\parallel = \langle u\rangle_\sigma - 5/2$, where $u = m_ev^2/2T_e$ and $\langle\cdot\rangle_\sigma$ is the average weighted by each speed's share of the conductivity.\cite{walsh_paper0}
	
	For a classical plasma in uniform $n_e$ the generation rate is $(1/e)(\partial \beta_\parallel / \partial Z) \nabla Z \times \nabla T_e$. The number of current-carriers rises faster on the high-$Z$ side of the ionization gradient. While the true electron density is uniform, the change in density of current-carriers from heating is not, due to e-e collisions\cite{walsh_paper0}.
	
	The kinetic suppression of composition Sadler then originates from the fact that current-carriers on both sides of the $Z$-gradient respond less to heating than $T_e^{\beta_{\parallel}}$ predicts. The suppression function becomes:

		\begin{equation}
	\beta_\parallel^k \equiv \beta_\parallel\Gamma_{nJ} + \Gamma_B - 1,
	\qquad
	\Gamma_Z = \frac{\partial_Z \beta_\parallel^k}{\partial_Z\beta_\parallel} \label{eq:Gam_Z}
	\end{equation}
	where $\beta_\parallel^k$ is the kinetic version of the coefficient.
	
	\begin{figure}[tb]
		\includegraphics[width=\linewidth]{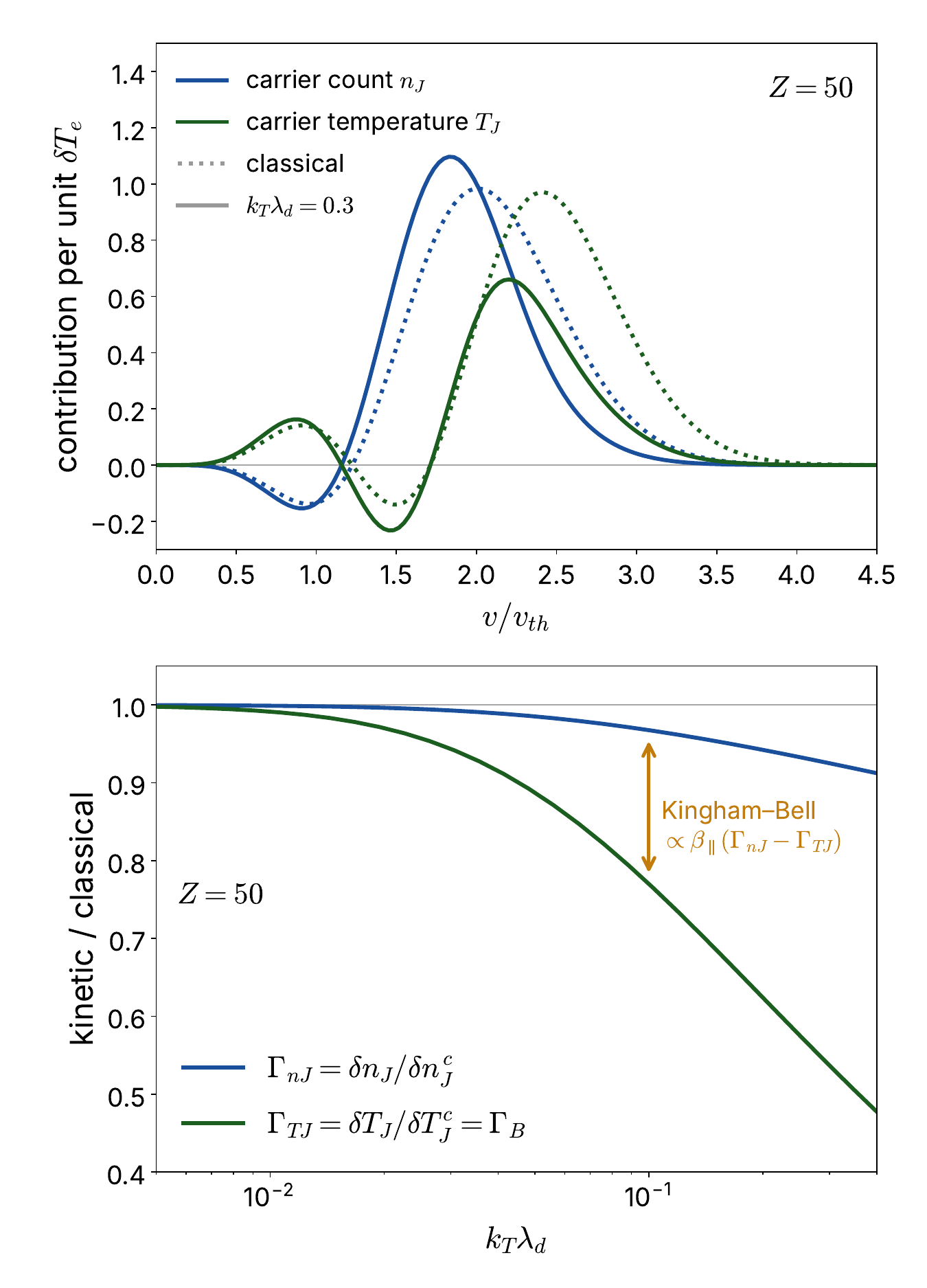}
		\caption{\label{fig:gen_carriers} How current-carrier temperature ($T_J$) and density ($n_J$) are affected by non-locality for $Z=50$. (a) Current-carrier moments of $f_0$ against electron velocity, with dotted lines for classical and solid lines for $k_T\lambda_d = 0.3$.  (b): Change in current-carrier temperature and density due to a temperature perturbation $\delta T$ against $k_T \lambda_d$. Quantities are normalized to their classical values ($\delta T_J/\delta T_J^c$, $\delta n_J/\delta n_J^c$). $T_J$, which depends on higher velocity electrons, is more affected by non-locality than $n_J$. Biermann suppression is the lag in current-carrier temperature: $\Gamma_B=\Gamma_{TJ}=\delta T_J/\delta T_J^c$. Composition Sadler suppression is $\Gamma_Z = \partial_Z(\beta_\parallel \Gamma_{nJ} + \Gamma_B) / \partial_Z \beta_\parallel$. Kingham-Bell emerges as $1-\Gamma_{KB} = \beta_\parallel(\Gamma_{nJ}-\Gamma_B)$.}
	\end{figure}
	
	\subsubsection{Kinetic origin of Kingham-Bell generation ($\nabla n_e=0$, $\nabla Z=0$) \label{sec:theory_kingham}}
	
	Kingham \& Bell\cite{kingham2002} observed a kinetic generation mechanism in VFP simulations with $\nabla n_e=0$ and $\nabla Z=0$. This process has no classical counterpart. 
	
	Classically, the number of current-carriers ($n_J$) and the current-carrier temperature ($T_J$) are locked together. In the kinetic regime this is not necessarily true. Consider a slowly varying background $T_e$ slope in $y$ ($\nabla \bar{T}_e$) and a hot perturbation $\delta T_e$ along $x$. Linearizing gives:
	\begin{equation}
		\nabla \ln n_J \times \nabla T_J \approx \nabla \ln \bar{n}_J \times \nabla \delta T_J + \nabla \delta \ln n_J \times \nabla \bar{T}_J
	\end{equation}

	Remember that the true electron density $n_e$ is uniform in this case; the number of current carriers $n_J$ is not. In the classical limit the two terms cancel. $n_J$ is set by $\langle v^3\rangle$, nearer the thermal speed, while $T_J$ also depends on $\langle v^5\rangle$, deeper in the depleted tail (see figure \ref{fig:gen_weights}). 
	
	Kinetically, the two properties respond differently to the perturbation: the carrier temperature lags by $\delta T_J = \Gamma_B\,\delta T_e$ and the carrier count by $\delta\ln n_J = \Gamma_{nJ}\,\beta_\parallel\,\delta T_e/\bar T_e$. Substituting (any $Z$):
	\begin{equation}
		\Big[ \frac{\partial \mathbf{B}}{\partial t} \Big]_{KB} = - \frac{\beta_{\parallel}}{e} (\Gamma_B - \Gamma_{nJ})\, \nabla \ln\bar{T}_e \times \nabla \delta T_e \chg{\label{eq:KB_theory}}
	\end{equation}
	This vanishes classically ($\Gamma_B=\Gamma_{nJ}=1$). Because the carrier temperature reads further into the depleted tail than the carrier count, $\Gamma_B<\Gamma_{nJ}$: a background temperature gradient acts like a density gradient pointing the other way. Where $\nabla\bar T_e$ is antiparallel to $\nabla\bar n_e$, as at an ablation front, it adds to the Biermann battery.

	Kingham \& Bell\cite{kingham2002} first identified this mechanism in 2D VFP simulations of a non-uniformly heated plasma with uniform density. A magnetic field grew to $\omega\tau\approx9$ within a picosecond, attributed to the separation of the $\langle v^3\rangle$ and $\langle v^5\rangle$ gradients; in the language of this paper, these are the carrier count and the carrier temperature. Their simulations used a Lorentz collision operator and $\mathbf{f}_1$ truncation at a single ionization ($Z=10$). The present work extends this in three ways. First, the full e-e collision operator and $Z=1$--100 are included; at low $Z$ the generation is weaker, through $\beta_\parallel(Z)$. Second, the steady generation rate is given as a function of $k\lambda_d$, in the same form as every other transport term, rather than an initial growth rate. Third, the mechanism is placed alongside the Biermann battery: when $\nabla\bar{T}_e$ is antiparallel to $\nabla n_e$, as at an ablation front, Kingham-Bell generation adds to the Biermann battery.
	
	\begin{figure*}[tb]
		\includegraphics[width=\textwidth]{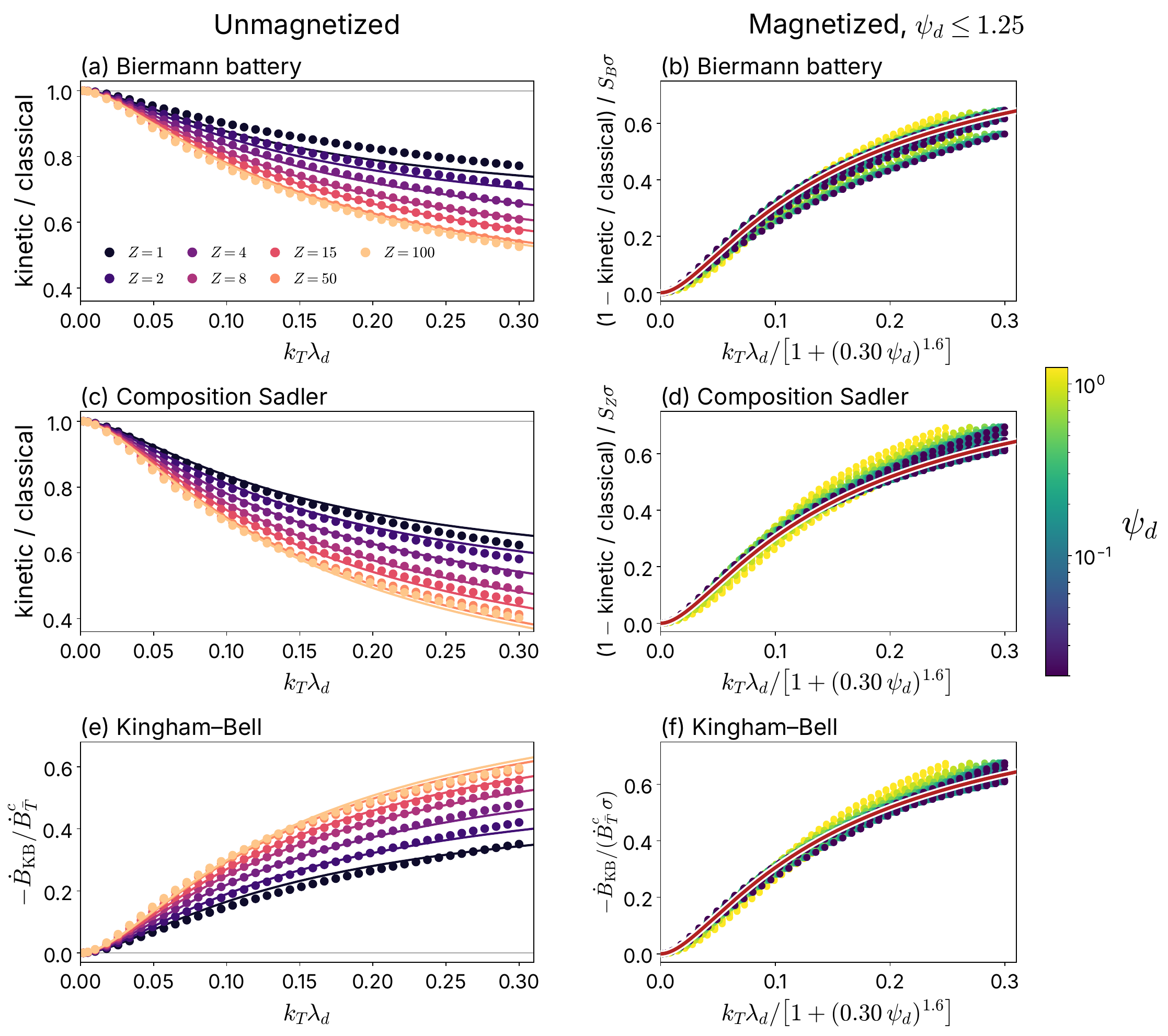}
		\caption{  \label{fig:gen_carrier_data} Left: Unmagnetized Biermann (a), composition Sadler (c) and Kingham-Bell generation (e) against $k_T \lambda_J$ for a range of $Z$. (a) and (c) are normalized to their classical value, while (e) uses $\dot{\mathbf B}_{KB}/\dot{\mathbf B}^c$ (where $\dot{\mathbf B}^c$ is classical Biermann rate) due to Kingham-Bell (KB) being zero in the classical regime. $k_T \lambda_{d}$ does not collapse the data, due to the amplitude $\sigma(Z)=(Z+1.54)/(Z+3.69)$. Right: magnetized versions of these processes, plotted to show the fit quality from equations \ref{eq:gamma_B_fit}, \ref{eq:gamma_Z_fit} and \ref{eq:gamma_KB_fit}. Dots are linear mode data, while the lines are the fits given in table \ref{tab:gamma}. The stress contributions have been removed for this figure. }  
	\end{figure*}
	
	\subsection{Kinetic Stress from $\underline{\underline{f}}_2$ \label{sec:theory_stress}}
	
	Ohm's law of equation \ref{eq:E_kinetic_origin} ignored electron distribution harmonics above $\mathbf{f}_1$. Recent
	work\cite{joglekarMagneticReconnectionPlasma2014,liu2017,walsh_paper2} found the kinetic stress originating in
	$\underline{\underline{f}}_2$ to play an important role in magnetic field advection. The shear stress ($T_{xy}$) was found to be particularly important in that case\cite{walsh_paper2}, allowing for $+y$-momentum traveling in $+x$ to stream past $-y$-momentum traveling in $-x$; without $T_{xy}$ (i.e. truncating at $\mathbf{f}_1$) these converging fluxes are forced to isotropize, requiring a large $E_y$ electric field to enforce the zero current condition\cite{walsh_paper2}. 
	
	For generation, it is instead the normal stress ($T_{xx}\ne T_{yy}$) that plays the important role. While the bulk stress is always small in the regime of this paper ($|T_{xx}-T_{yy}|/\delta T <0.1$), the stress in the distribution tail ($v\approx3.3v_{th}$) is typically more than an order of magnitude larger\cite{walsh_paper2}.

	A general form for the stress, which contains both the advective and generative components, is:
	\begin{equation}
		\mathbf{E}_{\Pi} = - \frac{\nabla \cdot \underline{\underline{\Pi}}_J}{e n_J} \label{eq:E_stress}
	\end{equation}
	where $\underline{\underline{\Pi}}_J$ is the direction-dependent part of the current-carrier pressure (the
	current-carrier-weighted moment of the $\underline{\underline{f}}_2$ term in the $\mathbf{f}_1$ equation, including the
	small $\mathbf{E}\cdot\underline{\underline{f}}_2$ contribution). $\mathbf{f}_1$ truncation necessarily gives $\mathbf{E}_{\Pi}=0$.
	
	The electron stress appears in the classical momentum equation too, but as viscosity driven by shear flow, which is negligible\cite{braginskiiTransportProcessesPlasma1965}. The kinetic stress is instead driven by the heat flow, scaling as $(k_T\lambda_{ei})^2$\cite{burnett1936}. Its local
	(small-$k_T$) form is the heat-flux viscosity of Liu \textit{et al.}\cite{liu2015}, who tabulated its magnetized
	coefficients and used it in a generalized Ohm's law\cite{liu2017}.
	
	For a temperature perturbation the extra pressure is along $\nabla T_e$. It does not change the total pressure, so any addition along the gradient is subtracted from the two directions across it:
	$\underline{\underline{\Pi}}_J = \Lambda_J (\hat{\mathbf{t}}\hat{\mathbf{t}} - \underline{\underline{I}}/3)$, with
	$\hat{\mathbf t}=\nabla T_e/|\nabla T_e|$. Along the gradient the field is a correction to the thermoelectric field,
	\begin{equation}
		e E_{\Pi,\parallel} = s\, T_e \nabla_\parallel \ln T_e , \label{eq:s}
	\end{equation}
	where $s(Z, k_T\lambda_d)$ is the stress as a fraction of $\nabla T_e$. It vanishes classically and grows as
	$(k_T\lambda_d)^2$; its fitted form is given in section \ref{sec:gen_stress}. Note that $s$ originates from the tail of the distribution, rather than from the bulk plasma.
	
	To first order, $s$ transforms $\beta_{\parallel}$ into $\beta_{\parallel}-s$, lowering the thermoelectric field. The stress is largest where the plasma is least collisional. Biermann generation exists because the same $\nabla T_e$ pushes harder per electron on the low-density side; the stress is also largest there (low density is less collisional) and acts against that push, so it suppresses Biermann. Composition Sadler exists because $\beta_\parallel$ is smaller on the low-$Z$ side; the stress is larger there (for constant $n_e$, low-$Z$ is less collisional), making $\beta_\parallel-s$ smaller still, so it enhances composition Sadler.
	
	In a uniform plasma $\mathbf E_\Pi$ is a gradient and generates nothing. Across a background gradient in $X$ ($n_e$ or
	$Z$) the stress is stronger on the less collisional side, and its sideways part, half the size of the part along
	$\nabla T_e$ and opposite in sign, no longer cancels. The curl is then
	\begin{equation}
		\frac{\dot{\mathbf{B}}_{\Pi}}{\dot{\mathbf{B}}_X^c} = \frac{3}{2}\frac{\partial s}{\partial \ln X} + c_X\, s \label{eq:curl_rule}
	\end{equation}
	with $\dot{\mathbf{B}}_X^c$ from equation \ref{eq:Bc_X} and the derivative taken at fixed $k_T$. The first term is how
	quickly the stress changes across the background gradient. The second is a smaller stress aligned exactly with
	$\nabla T_e$ giving $c_X = \frac{1}{2}\,\partial\ln n_J/\partial\ln X$, from the $1/n_J$ in equation \ref{eq:E_stress}. The background gradient adds a part that is not aligned; $c_X$ is fitted in section \ref{sec:gen_stress}. For
	Kingham-Bell both gradients are temperature gradients, the stress is not aligned with either, and it is fitted
	separately.
	
	The same stress field therefore has three roles: along $\nabla T_e$ it lowers the thermoelectric field, across
	$\nabla T_e$ it is the Nernst stress\cite{walsh_paper2}, and across a background gradient it
	generates magnetic field.
	
		\begin{figure*}[tb]
		\includegraphics[width=\textwidth]{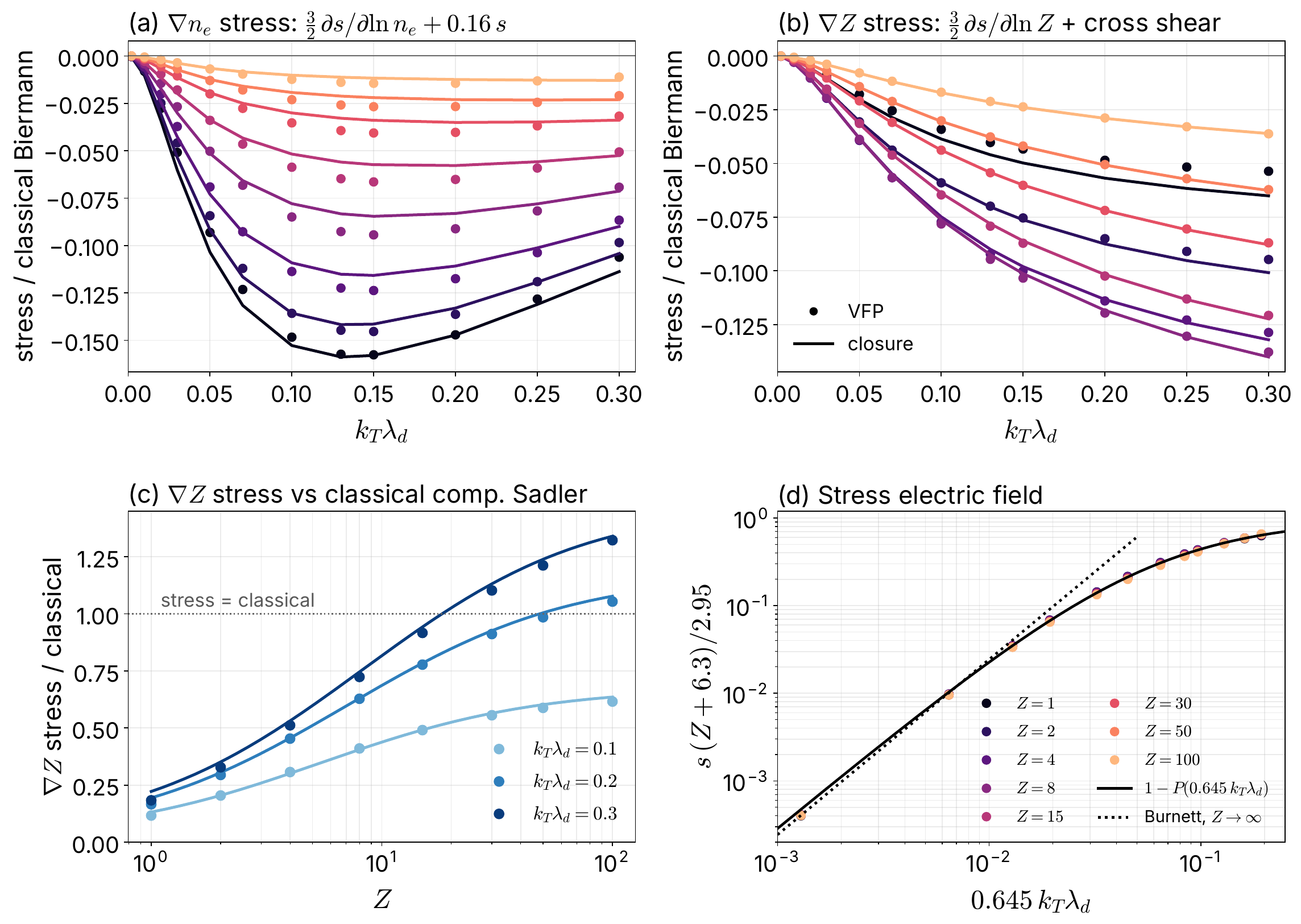}
		\caption{  \label{fig:gen_onestress} (a) The stress contribution to $\dot{\mathbf B}$ in a density gradient, normalized to the classical Biermann rate. Low $Z$ is more affected by the stress term. (b) Stress contribution to $\dot{\mathbf B}$ in an ionization gradient, normalized to $\dot{\mathbf B}^c_Z=-(T_e/e)\nabla\ln Z\times\nabla\ln T_e$  (c) The stress contribution at an ionization gradient grows with $Z$ and $k_T \lambda_{d}$, exceeding the composition sadler rate for $k_T \lambda_d=0.3$, $Z\ge19$. (d) full stress contribution to the electric field, with the line showing the fit presented in table \ref{tab:gamma}. Dashed line is the local Burnett stress\cite{burnett1936,liu2015}.}
	\end{figure*}

	\section{Simulations \label{sec:sims}}
	
	Kinetic inhibition of magnetic field generation is measured in the linear regime using two tools. Fits are made to a quasi-static linear mode solver\cite{walsh_paper1}, which allows a dense database of $>4\times 10^4$ simulations across ionization, magnetization and perturbation wavelength for each of the three background gradients required to drive generation ($\nabla n_e$, $\nabla Z$, $\nabla \bar{T}_e$). Time-dependent VFP simulations validate the quasi-static solver and provide simultaneous measurements for figure \ref{fig:closing_all_terms}. Both simulation types expand the electron distribution in spherical harmonics up to $\underline{\underline{f}}_6$ with the full e-e collision operator\cite{tzoufrasVlasovFokkerPlanck2011}, treating the ions as stationary. The importance of the full collision operator was discussed in earlier work \cite{walsh_paper0,walsh_paper1}; appendix \ref{sec:davies} here shows its impact on Biermann suppression.
	
	\subsection{Geometry \label{sec:geometry}}
	1D simulations are not adequate for this work, as field generation requires misaligned gradients. A temperature perturbation $\delta T\propto\cos(k_Ty)$ of small amplitude runs along $y$. A slowly varying background gradient in $n_e$ (Biermann), $Z$ at fixed $n_e$ (composition Sadler) or $T_e$ (Kingham-Bell) runs along $x$, with its own scale much longer than the mean free path. The response is linear in $\delta T$\cite{walsh_paper1}. The background gradients are slowly varying, which removes dependence on that length-scale; it is the perturbation length in $y$ that sets the wavenumber $k$. A uniform magnetic field along $z$ is imposed and held fixed; $\dot{\mathbf B}=-\nabla\times\mathbf E$ is measured but not fed back, so the electron response is isolated at a chosen magnetization.
	
	Only temperature perturbations carry kinetic corrections. A density perturbation at uniform temperature leaves $f_0$ Maxwellian, verified to $3\times10^{-4}$.
	
	\subsection{Linear mode solver \label{sec:eigen}}
	The solver of the previous papers\cite{walsh_paper1,walsh_paper2} is extended to first order in the background gradient. For each wavenumber the temperature perturbation is the slowest decaying mode, with the higher harmonics quasi-static and the electric field set by zero current. Each run returns the full field and its split into current-carrier, stress and gyration parts (equations~\ref{eq:Bgen_kin} and \ref{eq:E_stress}); the split is exact. 
	
	The database covers 27 values of $Z$ from 1 to 100, 40 values of $k_T\lambda_d$ from 0.002 to 0.3, and 13 values of $\psi_d$ from 0 to 1.25. Velocity space uses $v_{max}=8v_{th}$. The velocity-grid error is first order in $1/N_v$, so every point is computed at $N_v=128$ and 256 and extrapolated, $2X_{256}-X_{128}$. This changes the fitted constants by less than 2\% and the stress by at most 0.006 of the classical Biermann rate. Truncating at $\underline{\underline{f}}_6$ changes the generation rates by less than $3\times10^{-5}$ against $\underline{\underline{f}}_{10}$ for $k_T\lambda_d\le0.1$. Convergence in harmonics is documented for all terms in appendix \ref{sec:harmonic_trunc}.

	\subsection{Time-dependent VFP \label{sec:td}}
	Time-dependent VFP simulations\cite{sherlock2017,sherlockSuppressionBiermannBattery2020,shaffer2023} start from Maxwellian and are run until the field-generation rate is steady. A '1.5D' geometry is used, with the perturbation linearized about the prescribed background. The background gradient runs along $x$ (spatially resolved), while the perturbation runs along $y$ ($\delta T\propto\cos(ky)$). The background's own variation is kept local ($k_x\lambda\ll0.01$). Previous work in the series validated linearity in perturbation amplitude $\delta$\cite{walsh_paper1}, allowing this approach.
	
		\begin{figure}[tbp]
		\includegraphics[width=0.95\linewidth]{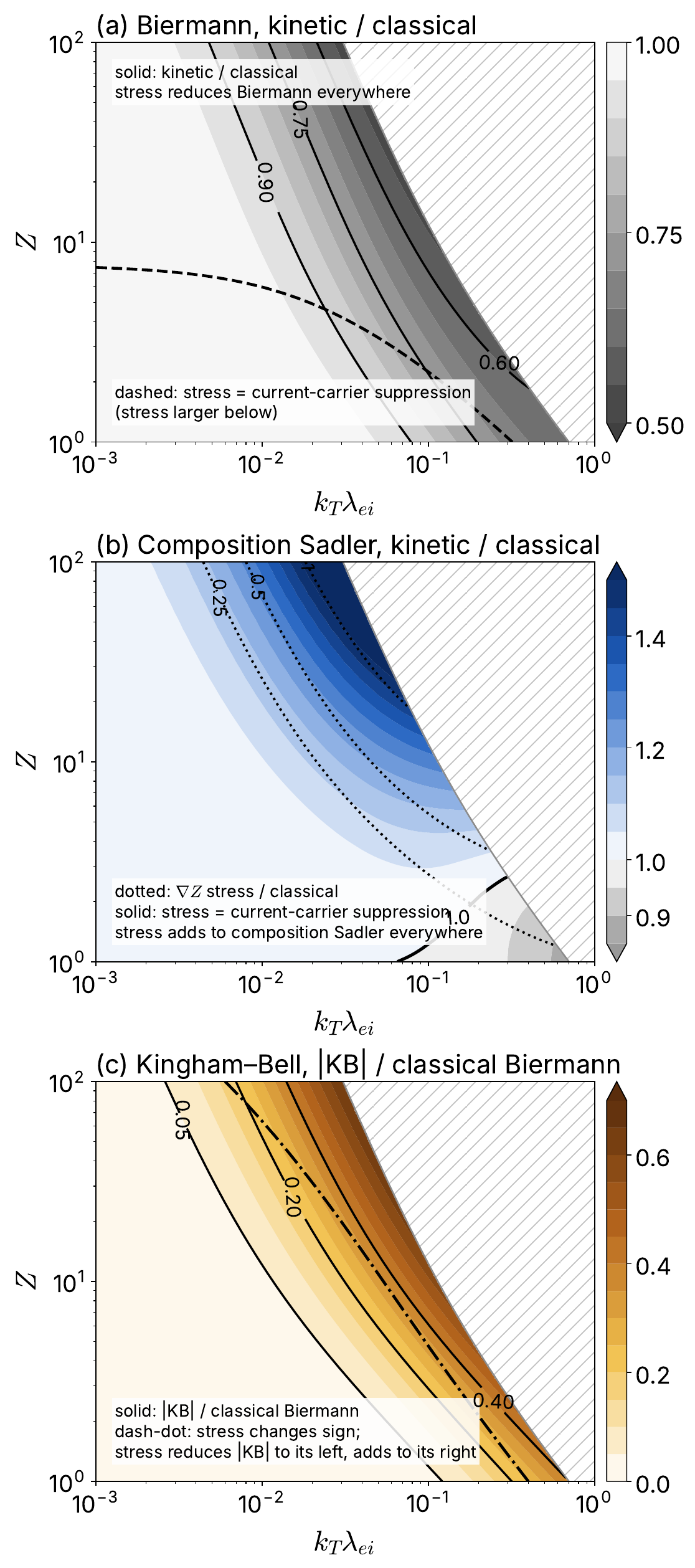}
		\caption{\label{fig:gen_regime} Suppression/enhancement of unmagnetized generation rates across ionization ($Z$) and non-locality ($k_T \lambda_{ei}$) with the stress contribution included. (a) Biermann battery, (b) composition Sadler, (c) Kingham-Bell. The striped region shows the regime outside the bounds of this paper. Dashed line in (a) shows where the stress term equals the current-carrier suppression (both lowering generation), with stress exceeding carrier suppression in the bottom left corner. Dotted lines in (b) show how much the stress term enhances composition Sadler, while the solid line shows where stress is equal to carrier suppression (above the line stress dominates). The dash-dot line in (c) shows where the stress changes sign, enhancing Kingham-Bell generation to the right. }
	\end{figure}
	
	Every transport process is read from the same simulation at the point of maximum background gradient: heat flow, Righi-Leduc, Nernst, cross-Nernst, and each generation mechanism. The runs use $Z=1$ and $Z=50$, spanning the range $\chi=0,0.4$ and $k_T\lambda_d=0.025,0.2$. These points are left out of all fits. The fits to linear mode data are compared with the time-dependent simulations in figure \ref{fig:closing_all_terms}, showing excellent agreement. More quantitatively, across 246 comparisons the error between approaches is 0.0018 RMS (0.006 max). Kinetic heat-flux and Nernst suppression are also measured in these 1.5D simulations and found to agree with earlier 1D results \cite{walsh_paper1,walsh_paper2}.

	\section{Generation by the current carriers \label{sec:gen_current_carriers}}
	
	This section characterizes the current-carrier generation mechanisms without modification by kinetic stress; section \ref{sec:gen_stress} then studies the stress terms in isolation, before showing the combined picture.
	
	Section \ref{sec:theory_currentcarriers} introduced the concept of current-carrier temperature ($T_J$) and density ($n_J$) in magnetic field generation. They are the plasma conditions that would be inferred if the only information available was from electrons carrying current. For any Maxwellian, $T_J=T_e$ and $n_J$ depends only on $n_e$ and $T_e$, which is why classical transport has forgotten the importance of the current-carrying electron population. 
	
	Figure \ref{fig:gen_carriers}(a) shows the current-carrier moments of the electron distribution at $Z=50$, for a classical plasma (dotted) and one with $k_T\lambda_{d}=0.3$ (solid). $T_J$ depends on higher energy electrons ($v\approx 2.4v_{th}$) than $n_J$ ($v\approx 2.0v_{th}$). The higher energy population is less collisional, making $T_J$ more affected by non-locality. The area under the kinetic curves normalized to the classical curves are $\Gamma_{nJ}=0.93$ and $ \Gamma_B = 0.54$.
	
	Figure \ref{fig:gen_carriers}(b) plots the ratio of kinetic current-carrier parameters to their classical counterparts against non-locality $k_T \lambda_d$ for an unmagnetized $Z=50$ plasma. The drop in temperature as seen by the current-carriers is responsible for the suppression of Biermann battery in kinetic conditions. The difference between the two curves is the driving mechanism for the Kingham-Bell mechanism.
	
	The suppression $\Gamma_X$ of each mechanism is measured from the linear mode database and plotted in figure \ref{fig:gen_carrier_data}. The plots on the left are unmagnetized: a) Biermann battery ($\Gamma_B$); c) composition Sadler ($\Gamma_Z$); e) Kingham-Bell ($\Gamma_{KB}$). Kingham-Bell (as well as the stress terms discussed in section \ref{sec:gen_stress}) is zero classically, so is normalized by
	\begin{equation}
		\dot{\mathbf B}^c_{X} \equiv -\frac{T_e}{e}\,\nabla\ln X\times\nabla\ln T_e , \label{eq:Bc_X}
	\end{equation}
	which is classical Biermann $\dot{\mathbf B}^c_{n}$ for $X=n_e$. Kingham-Bell uses $X=\bar T_e$.
	
	\ref{fig:gen_carrier_data}(a,c) shows Biermann falling to 50--80\% of its classical value by $k_T\lambda_d=0.3$, while composition Sadler falls to 40--70\%. Kingham-Bell rises from zero to 60\% of $\dot{\mathbf B}^c_{\bar T}$.
	
	Note that, unlike heat-flux\cite{walsh_paper1} and pure Nernst\cite{walsh_paper2}, the generation terms at different ionization do not collapse onto one curve when plotted against $k_T \lambda_{d}$. Their amplitudes still depend on $Z$, which is why the $S_X$ column is needed in table \ref{tab:gamma}.

	The data is fit for each process using the entire linear database:
	\begin{equation}
		\Gamma_B = 1 - 0.75\sigma(Z)\left[1 - P\!\left(\frac{k_T\lambda_J}{1+(0.30\,\psi_d)^{1.6}}\right)\right] \label{eq:gamma_B_fit}
	\end{equation}
	\begin{equation}
		\Gamma_Z = 1 - \sigma(Z)\left[1 - P\!\left(\frac{k_T\lambda_J}{1+(0.30\,\psi_d)^{1.6}}\right)\right] \label{eq:gamma_Z_fit}
	\end{equation}
	\begin{equation}
		\Gamma_{KB} = 1 - \sigma(Z)\left[1 - P\!\left(\frac{k_T\lambda_J}{1+(0.30\,\psi_d)^{1.6}}\right)\right] \label{eq:gamma_KB_fit}
	\end{equation}
	with $\lambda_J = 0.64\,\lambda_d$, $\sigma(Z) = (Z+1.54)/(Z+3.69)$. $P(K)$ (equation \ref{eq:P}) is the same suppression function as for heat-flux\cite{walsh_paper1}, Righi-Leduc, Nernst\cite{walsh_paper2} and cross-gradient Nernst.

	With the $Z$-dependent amplitude $\sigma(Z)$, the data falls approximately onto one curve for each process, as shown on the right-hand-side of figure \ref{fig:gen_carrier_data}. The three mechanisms share a length-scale $\lambda_J$ and a delocalization Hall parameter $0.30\psi_d$, with the same $Z$-dependence on the amplitude.

	\section{Generation by stress \label{sec:gen_stress}}
	This section isolates generation by the stress terms, splitting them into components by the background gradients, as was done for the carrier terms (section \ref{sec:gen_current_carriers}). The section ends by showing the combined effect for each background gradient.
	
	Figure \ref{fig:gen_onestress} plots magnetic field generation due to stress in simulations with no initial magnetic field. (a) is with a background $\nabla n_e$; (b) uses background $\nabla Z$. The points are from the linear mode database, while the lines are the fits presented later in this section. The stress is normalized to classical Biermann in (a) and to $\dot{\mathbf B}_Z^c$ in (b). For a background density gradient, the stress is stronger in magnitude at low $Z$. For a background ionization gradient, the stress magnitude peaks at moderate ionization.
	
	Figure \ref{fig:gen_onestress}(c) shows the $\nabla Z$ stress normalized to composition Sadler, plotted against ionization on the $x$-axis for 3 values of $k_T\lambda_{d}$. Higher $Z$ and higher non-locality increase the stress, with stress exceeding the classical composition Sadler rate for $k_T \lambda_{d}=0.3$,$Z\ge19$.
	
	The stress field of equation \ref{eq:s} is fitted by
	\begin{equation}
		s = \frac{2.95}{Z+6.3}\,(1-P(K)) \qquad K = 0.645\,k_T\lambda_d \label{eq:s_fit}
	\end{equation}
	The exact small-$k_T$ limit is
	\begin{equation}
		s \rightarrow C_\Pi (k_T\lambda_d)^2, \qquad C_\Pi = \frac{896/3}{Z+6.3} \label{eq:CPi}
	\end{equation}
	The fit is 21\% above this in the limit; fit quality at finite $k_T\lambda_d$ was prioritized over an exact limit where the magnitude tends to zero.
	
	The generation terms use the derivatives of $s$ at fixed $k_T$. With $\lambda_d\propto 1/n_e$ and $P' = dP/dK$:
	\begin{equation}
		\frac{\partial s}{\partial\ln n_e} = \frac{2.95\,KP'(K)}{Z+6.3} \label{eq:dsdn}
	\end{equation}
	\begin{equation}
		\frac{\partial s}{\partial\ln Z} = -\frac{Z}{Z+6.3}\left[s - \frac{1.475\,KP'(K)}{Z+4.45}\right] \label{eq:dsdZ}
	\end{equation}
	
	The stress generation across each background gradient is then
	\begin{align}
		\dot{\mathbf B}_{\Pi,n} &= \left[\tfrac32\,\frac{\partial s}{\partial\ln n_e} + 0.16\,s\right]\mathcal F\,\dot{\mathbf B}^c_{n} \label{eq:stress_n}\\
		\dot{\mathbf B}_{\Pi,Z} &= \Big[\tfrac32\,\frac{\partial s}{\partial\ln Z} + 0.47\,w\,s\,\big(1-(1+2.2\,w) \nonumber\\
		&\qquad\ \ \times (1-P(0.57\,k_T\lambda_d))\big)\Big]\mathcal F\,\dot{\mathbf B}^c_{Z} \label{eq:stress_Z}\\
		\dot{\mathbf B}_{\Pi,\bar T} &= -\Big[0.38\,s + 0.69\,\frac{\partial s}{\partial\ln n_e} \nonumber\\
		&\qquad\ \ + 0.26\,(1-P(0.1\,k_T\lambda_d))\Big]\dot{\mathbf B}^c_{\bar T} \label{eq:stress_T}
	\end{align}
	with $w = Z/(Z+12.2)$ and $\dot{\mathbf B}^c_X$ from equation \ref{eq:Bc_X}. The first term of equations \ref{eq:stress_n}
	and \ref{eq:stress_Z} is the curl rule of equation \ref{eq:curl_rule}; the remaining terms are fitted.
	
	Figure \ref{fig:gen_onestress}(d) shows the efficacy of this fit, with all ionizations collapsing onto one curve ($s(Z+6.3)/2.95$) when plotted against $0.645 k_T \lambda_d$. The small-$k$ Burnett stress is shown as a dotted line.
	
	In a magnetic field the stress depends on one Hall parameter,
	\begin{equation}
		\psi_s = \frac{20.8\,Z\chi/(Z+6.6)}{1+k_T\lambda_d/0.105} \label{eq:psi_s}
	\end{equation}
	The stress field rotates as
	\begin{equation}
		e E_\Pi = T_e\,s\,(1-i\psi_s)^{-1.7}\,\partial_\parallel\ln T_e \label{eq:s_mag}
	\end{equation}
	where the real part is along $\nabla T_e$ and the imaginary part is across it (the Nernst stress\cite{walsh_paper2}).
	The $\nabla n_e$ and $\nabla Z$ generation is reduced by
	\begin{equation}
		\mathcal F = {\rm Re}\,(1+i\psi_s)^{-2.64} \label{eq:stress_rot}
	\end{equation}
	with $\mathcal F = 1$ unmagnetized. The Kingham-Bell stress is not rotated.
	
	The fits hold for $k_T\lambda_d\le 0.3$ and $\psi_d\le 1.25$. The stress field is fitted to an RMS of 0.004 along
	$\nabla T_e$ and 0.006 across it, as a fraction of $T_e\nabla\ln T_e$.
	
	A varied picture of kinetic stress' role in field generation can be seen through combining the generation from current-carrier terms (section \ref{sec:gen_current_carriers}) with the stress terms (this section). Figure \ref{fig:gen_regime} contains three contour plots, one for each background gradient: (a) $\nabla n_e$ (Biermann-battery-like); (b) $\nabla Z$ (composition-Sadler-like); (c) $\nabla \bar{T}_e$ (Kingham-Bell-like). The colorbar shows measured suppression/enhancement from the linear mode database, against ionization ($Z$) and non-locality ($k_T \lambda_{ei}$). The striped region denotes the regime outside the database's range.
	
	Biermann battery is suppressed $<1$ everywhere by non-locality. The stress term always acts to reduce generation, but is only dominant over carrier non-locality for a low-$Z$, moderately-kinetic plasma. Composition Sadler is enhanced $>1$ even at moderate $Z$, increasing with $Z$ and non-locality up to the parameter space edge. The kinetic stress is greater in magnitude than the kinetic current-carrier suppression everywhere except the bottom right corner (low $Z$, high non-locality). For Kingham-Bell, the stress lowers generation to the left of the dash-dot line, increasing it to the right. Generation up to 70\% of classical Biermann is found, landing in the region where Biermann gets reduced most.

	\begin{figure*}[tb]
		\includegraphics[width=\textwidth]{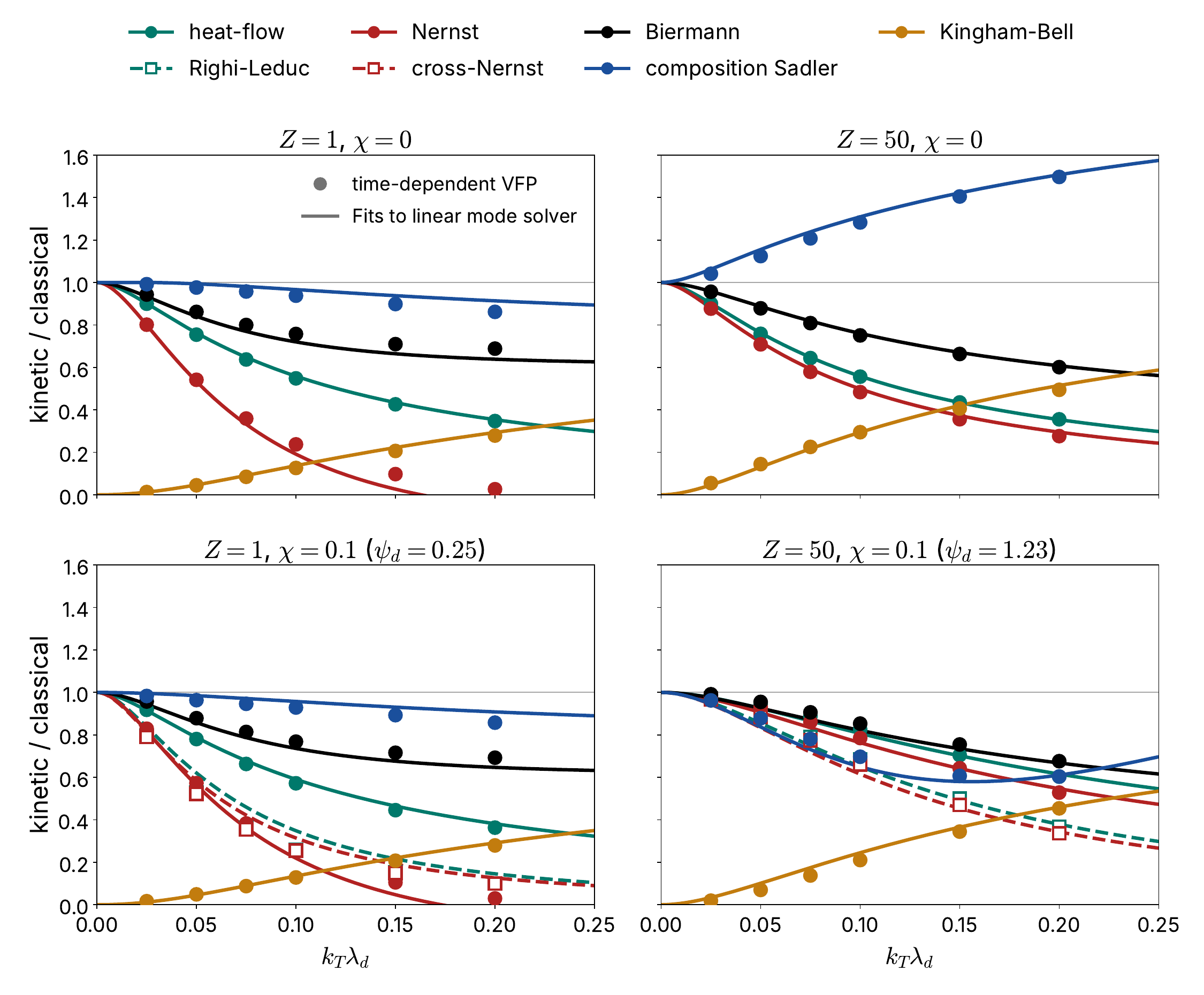}
		\caption{\label{fig:closing_all_terms} 
		Simultaneous measurements of all transport processes against severity of non-locality ($k\lambda_d$). Lines are the fits from table \ref{tab:gamma} including kinetic stress. Points are time-dependent 1.5D VFP simulations, which were not used in the fits.  (a): $Z=1$, unmagnetized. (b): $Z=50$, unmagnetized. (c): $Z=1$, $\chi=0.1$ ($\psi_d=0.25$). (d): $Z=50$, $\chi=0.1$ ($\psi_d=1.23$).  As Kingham-Bell generation (KB) is zero classically, it rises from zero and is normalized to the classical Biermann generation rate $-\dot{\mathbf B}_{KB}/\dot{\mathbf B}_B^c$. Magnetic flux generation is consistently less suppressed than the other transport processes, with composition Sadler enhanced by kinetic stress for unmagnetized $Z=50$. Nernst is more suppressed than the other mechanisms, particularly at low $Z$. } 
	\end{figure*}

	\section{Plasma transport by carrier electrons \label{sec:discussion}}

	\begin{figure}[tb]
		\includegraphics[width=\linewidth]{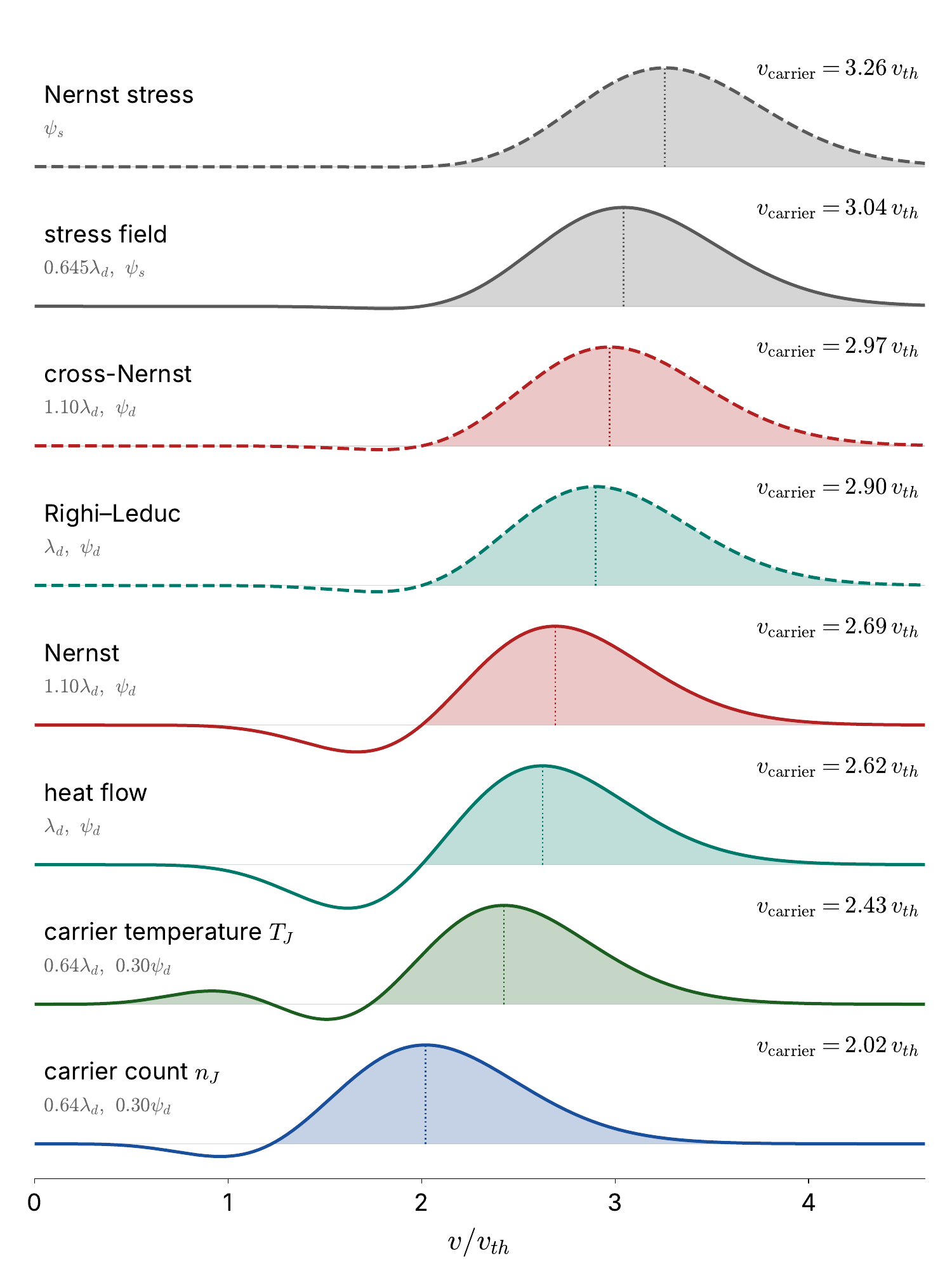}
		\caption{\label{fig:gen_ridge} Carrier population for each kinetic transport process, sorted such that the terms carried by faster electrons are to the top. Processes carried by faster electrons are broadly delocalized more easily (larger delocalization length) and are relocalized by magnetization more easily (larger delocalization Hall parameter).} 
	\end{figure}

	Plasma transport in the kinetic regime has been characterized using advanced methods (VFP with full electron-electron collision operator and converged electron distribution harmonics up to $f_6$) on a simple problem (linear perturbations in the marginally-kinetic regime) across 3 papers studying heat-flow\cite{walsh_paper1}, magnetic field advection\cite{walsh_paper2} and magnetic field generation (this work). More than $3\times10^5$ simulations were used to reach the following equations for the evolution of electron energy and magnetic field:
\begin{equation}
	\mathbf q_e = -\Gamma_{q\parallel}\kappa_\parallel\nabla_\parallel T_e - \Gamma_{q\bot}\kappa_\perp\nabla_\perp T_e
	- \Gamma_{q\bot}^2\,\kappa_\wedge\,\hat{\mathbf b}\times\nabla T_e \label{eq:q_kinetic}
\end{equation}
\begin{align}
	\frac{\partial\mathbf B}{\partial t} = \ &\nabla\times\Big[\big(\mathbf v_{plasma} - \Gamma_N^\ast\gamma_\perp\nabla_\perp T_e - \Gamma_N^2\gamma_\wedge\hat{\mathbf b}\times\nabla T_e\big)\times\mathbf B\Big] \nonumber\\
	&+ \frac{1}{e}\Big[\Gamma_B^\ast\,\nabla T_e\times\nabla\ln n_e + \Gamma_Z^\ast\,\nabla\beta_\parallel\times\nabla T_e \nonumber\\
	&\qquad + \Gamma_{\bar T}^\ast\,\nabla\ln\bar T_e\times\nabla T_e\Big] \label{eq:induction_kinetic}
\end{align}
with $\Gamma_{q\parallel}=P(k_T\lambda_d)$ and $\Gamma_{q\bot}=P(k_T\lambda_{d\perp})$ (equations \ref{eq:lam_dperp} and \ref{eq:P}). The asterisked terms have current-carrier suppression plus a stress contribution:
\begin{align}
	\Gamma_N^\ast &= \Gamma_N + \Delta\Gamma_N, \nonumber\\
	\Gamma_B^\ast &= \Gamma_B + \mathcal F\,\Pi_n, \nonumber\\
	\Gamma_Z^\ast &= \Gamma_Z - \mathcal F\,\Pi_Z\big/\left(Z\,\partial_Z\beta_\parallel\right), \nonumber\\
	\Gamma_{\bar T}^\ast &= 1 - \Gamma_{KB} + \Pi_{\bar T}, \label{eq:gamma_star}
\end{align}
where $\Pi_n$, $\Pi_Z$ and $\Pi_{\bar T}$ are the bracketed stress terms of equations \ref{eq:stress_n}--\ref{eq:stress_T}, and $\Gamma_N$ and $\Delta\Gamma_N$ the Nernst rows of table~\ref{tab:gamma}. $\mathcal F$ is the rotation of the stress from equation \ref{eq:stress_rot}. In the Kingham-Bell term $\nabla\bar T_e$ is the gradient of the background temperature, while $\nabla T_e$ is the gradient of the perturbation. Classically $\Gamma_N^\ast=\Gamma_B^\ast=\Gamma_Z^\ast=1$ and $\Gamma_{\bar T}^\ast=0$, recovering equation \ref{eq:mag_trans_new}.

For a uniform field, the suppression functions ($\Gamma_X$) can be calculated using table \ref{tab:gamma}, valid for $k_T\lambda_d\le0.3$ (for generation terms $\psi_d\le1.25$).

Equations \ref{eq:q_kinetic} and \ref{eq:induction_kinetic} are the pinnacle of this paper series, meeting the aim to tie together kinetic modification of all 8 plasma transport terms. Figure \ref{fig:closing_all_terms} demonstrates this objective by plotting the \emph{simultaneous measurement} of the kinetic suppression of heat-flux, Nernst, Biermann battery, composition Sadler and Kingham-Bell generation from time-dependent 1.5D VFP simulations (points). (a) and (b) are for unmagnetized plasmas with $Z=1$ and $Z=50$; (c) and (d) are then for $\chi=0.1$. The lines are the $\Gamma$ suppression functions of table \ref{tab:gamma}. 

Several observations stand out. Of the three famous transport mechanisms (heat-flux, Nernst advection and Biermann generation), Nernst is the most suppressed by kinetics, while Biermann battery is the least. Considering that Biermann magnetizes heat-fronts, while Nernst demagnetizes, this work points to an increased effect of self-generated magnetic fields when plasmas are in the kinetic regime. Righi-Leduc is more suppressed than perpendicular heat-flux (owing to being carried by higher velocity electrons\cite{walsh_paper0}), while cross-gradient-Nernst is \emph{less} suppressed than its perpendicular counterpart for $Z=1$; this is due to the significance of Nernst stress in these conditions\cite{walsh_paper2}, which does not affect cross-gradient-Nernst.

The fickle nature of composition Sadler should be noted as well: enhanced field generation for $Z=50$ when unmagnetized before becoming suppressed when magnetized to $\chi=0.1$ (more so than Biermann for $k_T \lambda_d\lesssim0.2$). 

The emergence of Kingham-Bell with rising $k_T \lambda_d$ also seems to have been neglected in the kinetic understanding of field generation\cite{sherlockSuppressionBiermannBattery2020,daviesNonlocalSuppressionBiermann2023}. Writing the background density as $n_e \propto \bar{T}_e ^\alpha$ allows a closer comparison with Biermann generation:
\begin{equation}
\frac{\dot{\mathbf{B} }_{\rm gen}}{\dot{\mathbf{B}} ^{c}_n} = \Gamma_B^\ast - \frac{\Gamma_{\bar T}^\ast}{\alpha}
\end{equation}
where $\dot{\mathbf{B}} ^{c}_n$ is the classical Biermann rate. For isobaric flow, such as an ablation front, Kingham-Bell adds to the Biermann field; figure \ref{fig:closing_all_terms} suggests what Biermann loses from kinetic suppression is regained by Kingham-Bell generation, leaving the total approximately constant (for $Z\gtrsim10$ and $k_T\lambda_d=0.1$--$0.3$, the net is $0.99$--$1.17$). For parallel $\nabla \bar{T}_e$ and $\nabla n_e$ (in an ablated corona, for example), $\alpha>0$; Kingham-Bell reinforces Biermann suppression. For a kinetic plasma with background temperature varying much faster than density ($|\alpha|\ll1$), Kingham-Bell is expected to dominate over Biermann battery.

Heat-flow flux limiters\cite{malone1975}, which are ubiquitous in fluid modeling of plasmas, make the substitution $k_T \lambda \to \lambda/(f L_T)$, where $L_T=T_e/|\nabla T_e|$ and $f$ is a tunable parameter\cite{jones2017,farmerHeatTransportModeling2018,doi:10.1063/5.0005776,farmer2021}. While most codes use a hard cap on heat-flow using $\lambda=\lambda_{ei}$ (i.e. $\kappa_\parallel ^c \le f_m \frac{L_T}{\lei}$), this work advocates for the smooth $\Gamma_q$ suppression function with the delocalization length $\lambda_d$.\cite{walsh_paper1,walshKineticCorrectionsHeatflow2024} Equations \ref{eq:q_kinetic} and \ref{eq:induction_kinetic} then allow for one tunable $f$ for all extended-MHD transport. For more discussion on how the value of $f$ depends on linear mode amplitude, see prior work in this series\cite{walsh_paper1}.

However, local flux limiters still have their limitations. On top of the usual suspects (such as having a tunable parameter, and not capturing long-range pre-heat), a flux limiter cannot model Kingham-Bell generation, which requires separating out the temperature background ($\bar{T}_e$) from the perturbation ($\delta T$). Such separation is not possible in a typical extended-MHD code, with cells that only know the conditions of their neighbors. In addition, the stress terms require consideration of the magnetic field profile, with the relative phase of the temperature and magnetic field perturbation significantly affecting the field advection\cite{walsh_paper2}.

When the magnetic field varies on scales comparable to $\lambda_d$, additional terms appear: the Nernst stress ($\Delta\Gamma_N$) depends on magnetic field wavelength;\cite{walsh_paper2} as does Righi-Leduc\cite{walsh_paper2}; and the stress generates field across $\nabla|\mathbf B|$. These depend on two wavenumbers and are better suited to a Fourier-space model than to local models such as extended-MHD; their closure is left to future work.

The concept that sustained this series is the importance of the 'carrier electrons': the electron population responsible for each transport process\cite{walsh_paper0}. Figure \ref{fig:gen_ridge} completes the loop, plotting the carrier population for each process as a function of $v/v_{th}$, ordered by fastest (top) to slowest (bottom). The processes with faster carrier electrons are delocalized more easily (longer length-scale) but are also relocalized more easily (higher delocalization Hall parameter). The cross-terms (cross-Nernst and Righi-Leduc) seem out of place, before remembering that they undergo suppression two-fold ($\Gamma_X^2$, see table \ref{tab:gamma}). Faster electrons are indeed more non-local, and are more easily magnetized. 

The stress terms are the exception. They have no classical counterpart, growing from zero with non-locality, and their length ($0.645\lambda_d$) sets where that growth saturates rather than where a classical process is lost. They are carried by the fastest electrons in the distribution ($\approx3v_{th}$), consistent with being relocalized most easily ($\psi_s>\psi_d$).

\section{Conclusion \label{sec:conclusion}} 

This paper has studied the delocalization of magnetic field generation in the moderately-kinetic linear regime, attributed to the temperature and density of the current-carrying electrons ($T_J$, $n_J$) lagging that of the bulk plasma ($T_e$, $n_e$). Biermann battery falls to 50\%--75\% of classical by $k_T \lambda_d=0.3$, while the emergence of Kingham-Bell generation (which has no classical counterpart) cancels Biermann suppression at isobaric fronts. Composition Sadler generation at ionization gradients is found to be enhanced for $Z\gtrsim 3$ due to kinetic stress. 

This paper series began by re-examining the classical coefficients, asserting that all transport comes from each process reading a different part of the electron distribution. With the current-carrier portions of the generation terms isolated (i.e. removing kinetic stress), field generation follows the suppression function for heat-flow\cite{walsh_paper1} with reduced delocalization length ($\lambda_J=0.64\lambda_d$) and reduced delocalization Hall parameter ($0.30\psi_d$); this is due to the current-carrier temperature and density relying on slower electron populations ($\approx2.43v_{th}$ and $2.02v_{th}$) than the heat-flow ($2.62v_{th}$). Kinetic theory can now relate the magnitude of each transport process to one another, albeit in the linear regime.

This work also highlights the importance of a full electron-electron collision model, as well as the retention of higher electron distribution harmonics up to \emph{at least} $\underline{\underline{f}}_2$. The former reduces non-locality by increasing the collisionality of faster electrons (see appendix \ref{sec:davies}), while the latter is fundamental to allowing fluxes to stream past one another\cite{walsh_paper2}.

By focusing on marginally non-local ($k\lambda_d\lesssim0.3$) linear ($\delta T_e \ll T_e$) plasmas, this study was able to thoroughly characterize transport suppression through three variables: ionization, magnetization, and $k\lambda_d$. Future work should extend this scope, studying highly-nonlocal and/or non-linear transport. While this may be a daunting project, the author would like to highlight their pleasant surprise in the simplicity of results found so far. Appendix \ref{sec:davies} provides a brief glimpse into how highly non-local plasmas behave, with Biermann generation \emph{relocalizing} for $k_T \lambda_d>0.4$. Future work should aim to reach the regime where the bulk plasma anisotropy is significant, rather than just the tail electrons characterized in the present study\cite{walsh_paper2}.

To make these results more applicable to direct-drive ablation fronts or hohlraum laser absorption, linear theory should be extended to incorporate Langdon heating from high laser intensities, which drives the equilibrium distribution super-Gaussian\cite{ridgersMagneticFieldsNonLocal2008,shaffer2023}. The carrier-electron framework seems appropriate for this endeavor. 

While the present work will have broadest reach in upgrading flux limiters in fluid models of plasma transport, it is also well-suited to providing the foundation for a multi-group electron transport model, such as the model of Schurtz-Nicola\"{i}-Busquet (SNB)\cite{schurtz2000}. While traditional SNB does a reasonable job of reproducing non-local unmagnetized heat-flux\cite{brodrickTestingNonlocalModels2017,sherlock2017}, it lacks extension to magnetized transport. The innate link between terms presented in this work is the crucial first step to such a model. The delocalization length (and delocalization Hall parameter) of this work may also help in setting the energy group blurring lengths.

\begin{acknowledgments}
	This work was performed under the auspices of the U.S. Department of Energy by Lawrence Livermore National Laboratory under Contract DE-AC52-07NA27344. 
	
	This document was prepared as an account of work sponsored by an agency of the United States government. Neither the United States government nor Lawrence Livermore National Security, LLC, nor any of their employees makes any warranty, expressed or implied, or assumes any legal liability or responsibility for the accuracy, completeness, or usefulness of any information, apparatus, product, or process disclosed, or represents that its use would not infringe privately owned rights. Reference herein to any specific commercial product, process, or service by trade name, trademark, manufacturer, or otherwise does not necessarily constitute or imply its endorsement, recommendation, or favoring by the United States government or Lawrence Livermore National Security, LLC. The views and opinions of authors expressed herein do not necessarily state or reflect those of the United States government or Lawrence Livermore National Security, LLC, and shall not be used for advertising or product endorsement purposes.
	
	The data that support the findings of this study are available from the corresponding author upon reasonable request. 
\end{acknowledgments}

\appendix


\section{Harmonic truncation \label{sec:harmonic_trunc}}
Table~\ref{tab:Lladder} gives the error from truncating the harmonic expansion at $f_L$, measured against $f_8$ with the linear mode solver. The heat flow, Nernst and Biermann rates are converged to within $10^{-3}$ by $f_4$ and $10^{-5}$ by $f_6$ for $k_T\lambda_d\le0.1$. The cross terms (Righi-Leduc and cross-Nernst) and composition Sadler converge more slowly. Truncation at $\mathbf f_1$ misses the composition Sadler enhancement entirely, an error of order unity, because that enhancement is the stress from $\underline{\underline{f}}_2$. All results in this paper use $f_6$, unless stated; spot checks against $f_{10}$ agree to within $3\times10^{-5}$ for $k_T\lambda_d\le0.1$.

\begin{table*}
	\caption{Harmonic truncation error: the largest difference between truncating the linear solver at $f_L$ and $f_8$, for the range $Z = 1$--$100$ and $\chi = 0$--$3$, for $k_T\lambda_d \le 0.1$ and $\le 0.28$. Every quantity is the full kinetic value normalized to classical. Truncation at $\mathbf f_1$ misses composition Sadler enhancement ($\dot{\mathbf B}_{\nabla Z}$) entirely.}
	\label{tab:Lladder}
	\footnotesize
	\setlength{\tabcolsep}{1.5pt}
	\begin{ruledtabular}
		\begin{tabular}{l@{\hspace{3pt}}|@{\hspace{3pt}}cccccc@{\hspace{3pt}}|@{\hspace{3pt}}cccccc}
			& \multicolumn{6}{c}{$k\lambda_d\le 0.1$} & \multicolumn{6}{c}{$k\lambda_d\le 0.28$} \\
			\cline{2-13}
			& $f_1$ & $f_2$ & $f_3$ & $f_4$ & $f_5$ & $f_6$ & $f_1$ & $f_2$ & $f_3$ & $f_4$ & $f_5$ & $f_6$ \\
			\hline
			$\mathbf{q}_\bot$ & $6.4\!\times\!10^{-2}$ & $1.1\!\times\!10^{-2}$ & $1.3\!\times\!10^{-3}$ & $1.4\!\times\!10^{-4}$ & $1.5\!\times\!10^{-5}$ & $4.4\!\times\!10^{-6}$ & $6.4\!\times\!10^{-2}$ & $2.0\!\times\!10^{-2}$ & $3.6\!\times\!10^{-3}$ & $8.2\!\times\!10^{-4}$ & $1.4\!\times\!10^{-4}$ & $2.2\!\times\!10^{-5}$ \\
			$\mathbf{q}_\wedge$ & $2.2\!\times\!10^{-1}$ & $7.2\!\times\!10^{-2}$ & $1.4\!\times\!10^{-2}$ & $1.7\!\times\!10^{-3}$ & $1.2\!\times\!10^{-4}$ & $2.7\!\times\!10^{-6}$ & $2.4\!\times\!10^{-1}$ & $9.5\!\times\!10^{-2}$ & $3.7\!\times\!10^{-2}$ & $1.3\!\times\!10^{-2}$ & $3.5\!\times\!10^{-3}$ & $7.8\!\times\!10^{-4}$ \\
			$\mathbf{v}_{N\perp}$ & $6.3\!\times\!10^{-2}$ & $1.2\!\times\!10^{-2}$ & $1.4\!\times\!10^{-3}$ & $1.6\!\times\!10^{-4}$ & $2.0\!\times\!10^{-5}$ & $1.0\!\times\!10^{-5}$ & $6.9\!\times\!10^{-2}$ & $1.7\!\times\!10^{-2}$ & $3.1\!\times\!10^{-3}$ & $7.1\!\times\!10^{-4}$ & $1.2\!\times\!10^{-4}$ & $1.7\!\times\!10^{-5}$ \\
			$\mathbf{v}_{N\wedge}$ & $2.3\!\times\!10^{-1}$ & $7.5\!\times\!10^{-2}$ & $1.5\!\times\!10^{-2}$ & $1.9\!\times\!10^{-3}$ & $1.5\!\times\!10^{-4}$ & $3.7\!\times\!10^{-6}$ & $2.4\!\times\!10^{-1}$ & $9.2\!\times\!10^{-2}$ & $3.6\!\times\!10^{-2}$ & $1.3\!\times\!10^{-2}$ & $3.7\!\times\!10^{-3}$ & $8.6\!\times\!10^{-4}$ \\[3pt]
			$\dot{\mathbf B}_{\nabla n_e}$ & $1.2\!\times\!10^{-1}$ & $4.1\!\times\!10^{-2}$ & $7.5\!\times\!10^{-3}$ & $1.1\!\times\!10^{-3}$ & $1.4\!\times\!10^{-4}$ & $1.6\!\times\!10^{-5}$ & $1.2\!\times\!10^{-1}$ & $1.3\!\times\!10^{-1}$ & $4.7\!\times\!10^{-2}$ & $1.6\!\times\!10^{-2}$ & $4.3\!\times\!10^{-3}$ & $1.0\!\times\!10^{-3}$ \\
			$\dot{\mathbf B}_{\nabla Z}$ & $5.6\!\times\!10^{-1}$ & $6.5\!\times\!10^{-2}$ & $1.2\!\times\!10^{-2}$ & $1.8\!\times\!10^{-3}$ & $2.4\!\times\!10^{-4}$ & $2.8\!\times\!10^{-5}$ & $1.3$ & $3.0\!\times\!10^{-1}$ & $1.1\!\times\!10^{-1}$ & $3.4\!\times\!10^{-2}$ & $9.3\!\times\!10^{-3}$ & $2.1\!\times\!10^{-3}$ \\
			$\dot{\mathbf B}_{\nabla \bar{T}}$ & $5.9\!\times\!10^{-2}$ & $3.1\!\times\!10^{-2}$ & $4.8\!\times\!10^{-3}$ & $5.8\!\times\!10^{-4}$ & $5.5\!\times\!10^{-5}$ & $4.1\!\times\!10^{-6}$ & $7.0\!\times\!10^{-2}$ & $1.9\!\times\!10^{-1}$ & $4.8\!\times\!10^{-2}$ & $1.3\!\times\!10^{-2}$ & $1.9\!\times\!10^{-3}$ & $2.6\!\times\!10^{-4}$ \\
		\end{tabular}
	\end{ruledtabular}
\end{table*}

\begin{figure}[tb]
	\includegraphics[width=\linewidth]{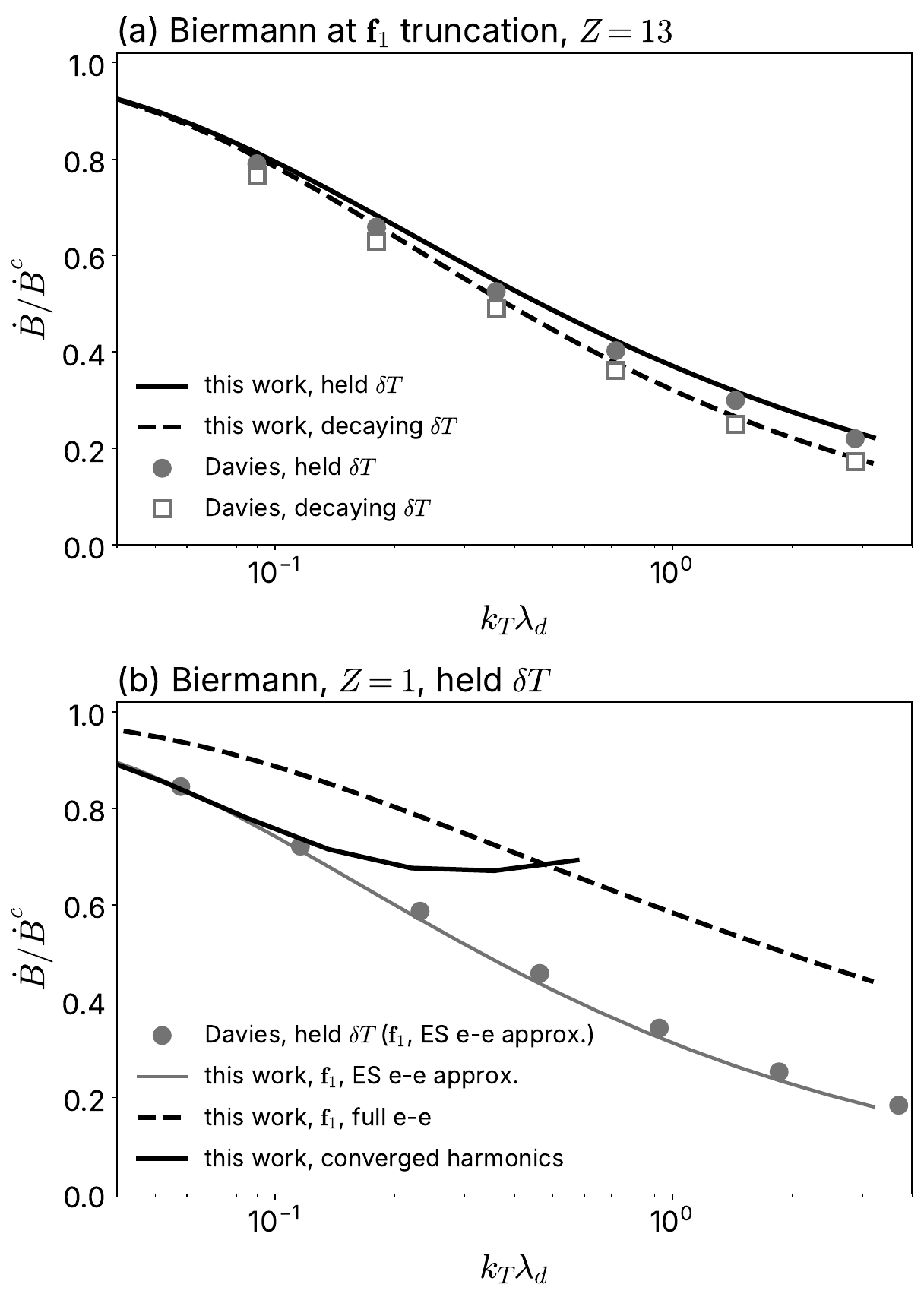}
	\caption{\label{fig:gen_davies} Comparison of Biermann suppression from the linear mode solver of this paper to the time-dependent VFP data published by Davies\cite{daviesNonlocalSuppressionBiermann2023}. (a) $Z=13$ with all calculations using harmonic truncation at $\mathbf{f}_1$. Simulations are shown with the temperature profile is allowed to decay (dashed line=this work; squares=Davies) and with the temperature profile held constant (solid line=this work; circles=Davies). (b) for $Z=1$ using a held temperature profile. The solid grey line is the linear solver of this work truncated at $\mathbf{f}_1$ using the same collision approximation as Davies\cite{epperleinPracticalNonlocalModel1991}. The dashed line includes the full collision operator, which localizes Biermann generation. The solid black line then uses converged harmonics up to $f_{10}$ as well as the complete collision operator. }
\end{figure}

\section{Comparison to published literature \label{sec:davies}}

Sherlock \& Bissell (SB) were the first to report and characterize Biermann battery suppression\cite{sherlockSuppressionBiermannBattery2020} on non-linear ablation fronts. Davies extended that work by simulating linear 2D $T_e$ and $n_e$ perturbations for two ionizations and across magnetization\cite{daviesNonlocalSuppressionBiermann2023}. The work of this paper is another extension, with these improvements:
\begin{itemize}
	\item The use of a full collision operator\cite{tzoufrasVlasovFokkerPlanck2011,walsh_paper1}, while Davies and SB used the Epperlein-Short approximation for e-e collisions\cite{daviesNonlocalSuppressionBiermann2023,epperlein1991}.
	\item The use of a converged harmonic expansion (up to $f_6$ in the main text and up to $f_{10}$ in this appendix), while Davies and SB truncated at $\mathbf{f}_1$.
	\item Davies and SB used only $\nabla n_e$ in the second dimension, missing suppression of composition Sadler and Kingham-Bell generation. This work uses $\nabla Z$ and $\nabla \bar{T}_e$.
\end{itemize} 

Two sets of simulations will be used for comparison. First, agreement with Davies' calculations under the same conditions validate the tools in this paper. The second demonstrates how improved treatment of the distribution and collision operator are significant.

Figure \ref{fig:gen_davies}(a) is the validation, with the points being $Z=13$ data calculated by Davies. Circles use a 'held' temperature perturbation, where the temperature profile is continuously driven to maintain the hot temperature. Squares use a 'decaying' profile, where the temperature is allowed to relax. Results using the linear mode solver of this paper are plotted as lines, with solid for held $\delta T$ and dashed for decaying $\delta T$; both lines truncate the distribution at $\mathbf{f}_1$ and adopt the collision approximation.

Figure \ref{fig:gen_davies}(b) shows Davies' data for $Z=1$ with the temperature perturbation held at the initialized value. The solid grey line is the linear mode solver of this work, with the same collision approximation\cite{epperleinPracticalNonlocalModel1991} as used by Davies, as well as $\mathbf{f}_1$ truncation. The dashed line shows the relocalizing effect of a full e-e collision operator at this ionization. The solid black line then uses a converged harmonic expansion, as well as the full collision operator. 

Note that Davies' calculations extended to much larger non-locality than the main body of this work, with $k_T \lambda_{d}$ up to $7$. The $f_6$ truncation used throughout this manuscript was found to be unconverged at these values of $k_T \lambda_{d}$, which led to extending the harmonic expansion. $f_{10}$ is converged.

Interestingly, the solid black line shows Biermann battery relocalizing for $k_T \lambda_d > 0.4$. This will be the subject of future work extending the regime of applicability of the linear theory in this paper. The key point of the appendix stands: harmonic convergence and electron collision physics affect kinetic field generation.

	\bibliographystyle{aipnum4-2}
	\bibliography{library}

@article{arran2025,
  title = {The Accuracy of Multi-Group Models for Nonlocal Electron Transport in Magnetized Plasmas},
  author = {Arran, C. and Holec, M. and Sherlock, M. and Marinak, M. and Ridgers, C. P.},
  year = 2025,
  month = apr,
  journal = {Physics of Plasmas},
  volume = {32},
  number = {4},
  pages = {043908},
  issn = {1070-664X},
  doi = {10.1063/5.0250147},
  urldate = {2026-08-06}
}

@article{bell1981,
  title = {Elecron {{Energy Transport}} in {{Steep Temperature Gradients}} in {{Laser-Produced Plasmas}}},
  author = {Bell, A. R. and Evans, R. G. and Nicholas, D. J.},
  year = 1981,
  month = jan,
  journal = {Physical Review Letters},
  volume = {46},
  number = {4},
  pages = {243--246},
  publisher = {American Physical Society},
  doi = {10.1103/PhysRevLett.46.243}
}

@inproceedings{braginskiiTransportProcessesPlasma1965,
  title = {Transport {{Processes}} in a {{Plasma}}},
  booktitle = {Reviews of {{Plasma Physics}}},
  author = {Braginskii, S. I.},
  year = 1965,
  volume = {1},
  pages = {205--205}
}

@article{brodrickIncorporatingKineticEffects2018,
  title = {Incorporating Kinetic Effects on {{Nernst}} Advection in Inertial Fusion Simulations},
  author = {Brodrick, J P and Sherlock, M and Farmer, W A and Joglekar, A S and Barrois, R and Wengraf, J and Bissell, J J and Kingham, R J and Sorbo, D Del and Read, M P and Ridgers, C P},
  year = 2018,
  month = jun,
  journal = {Plasma Physics and Controlled Fusion},
  volume = {60},
  number = {8},
  pages = {084009},
  publisher = {IOP Publishing},
  issn = {0741-3335},
  doi = {10.1088/1361-6587/aaca0b}
}

@article{brodrickTestingNonlocalModels2017,
  title = {Testing Nonlocal Models of Electron Thermal Conduction for Magnetic and Inertial Confinement Fusion Applications},
  author = {Brodrick, J. P. and Kingham, R. J. and Marinak, M. M. and Patel, M. V. and Chankin, A. V. and Omotani, J. T. and Umansky, M. V. and Del Sorbo, D. and Dudson, B. and Parker, J. T. and Kerbel, G. D. and Sherlock, M. and Ridgers, C. P.},
  year = 2017,
  month = sep,
  journal = {Physics of Plasmas},
  volume = {24},
  number = {9},
  pages = {092309--092309},
  doi = {10.1063/1.5001079}
}

@article{burnett1936,
  title = {The {{Distribution}} of {{Molecular Velocities}} and the {{Mean Motion}} in a {{Non-Uniform Gas}}},
  author = {Burnett, D.},
  year = 1936,
  month = jan,
  journal = {Proceedings of the London Mathematical Society},
  volume = {s2-40},
  number = {1},
  pages = {382--435},
  publisher = {John Wiley \& Sons, Ltd},
  issn = {0024-6115},
  doi = {10.1112/plms/s2-40.1.382},
  urldate = {2026-11-21}
}

@misc{campbell2021measuring,
  title = {Measuring Magnetic Flux Suppression in High-Power Laser-Plasma Interactions},
  author = {Campbell, P. T. and Walsh, C. A. and Russell, B. K. and Chittenden, J. P. and Crilly, A. and Fiksel, G. and Gao, L. and Igumenshchev, I. V. and Nilson, P. M. and Thomas, A. G. R. and Krushelnick, K. and Willingale, L.},
  year = 2021,
  eprint = {2107.12864},
  primaryclass = {physics.plasm-ph},
  archiveprefix = {arXiv}
}

@article{catto2005,
  title = {A New, Explicitly Collisional Contribution to the Gyroviscosity and the Radial Electric Field in a Collisional Tokamak},
  author = {Catto, P. J. and Simakov, A. N.},
  year = 2005,
  month = nov,
  journal = {Physics of Plasmas},
  volume = {12},
  number = {11},
  pages = {114503},
  issn = {1070-664X},
  doi = {10.1063/1.2136355},
  urldate = {2026-11-22}
}

@article{daviesNonlocalSuppressionBiermann2023,
  title = {Nonlocal Suppression of {{Biermann}} Battery Magnetic-Field Generation for Arbitrary Atomic Numbers and Magnetization},
  author = {Davies, J. R.},
  year = 2023,
  month = jul,
  journal = {Physics of Plasmas},
  volume = {30},
  number = {7},
  pages = {072701},
  issn = {1070-664X},
  doi = {10.1063/5.0152530},
  urldate = {2023-09-19}
}

@article{doi:10.1063/5.0005776,
  title = {Validation of Heat Transport Modeling Using Directly Driven Beryllium Spheres},
  author = {Farmer, W. A. and Bruulsema, C. and Swadling, G. F. and Sherlock, M. W. and Rosen, M. D. and Rozmus, W. and Edgell, D. H. and Katz, J. and Pollock, B. B. and Ross, J. S.},
  year = 2020,
  journal = {Physics of Plasmas},
  volume = {27},
  number = {8},
  eprint = {https://doi.org/10.1063/5.0005776},
  pages = {082701},
  doi = {10.1063/5.0005776}
}

@article{epperlein1991,
  title = {A Practical Nonlocal Model for Electron Heat Transport in Laser Plasmas},
  author = {Epperlein, E. M. and Short, R. W.},
  year = 1991,
  month = nov,
  journal = {Physics of Fluids B: Plasma Physics},
  volume = {3},
  number = {11},
  pages = {3092--3098},
  issn = {0899-8221},
  doi = {10.1063/1.859789},
  urldate = {2026-08-06}
}

@article{epperleinPlasmaTransportCoefficients1986,
  title = {Plasma Transport Coefficients in a Magnetic Field by Direct Numerical Solution of the {{Fokker}}--{{Planck}} Equation},
  author = {Epperlein, E. M. and Haines, M. G.},
  year = 1986,
  journal = {Physics of Fluids},
  volume = {29},
  number = {4},
  doi = {10.1063/1.865901}
}

@article{epperleinPracticalNonlocalModel1991,
  title = {A Practical Nonlocal Model for Electron Heat Transport in Laser Plasmas},
  author = {Epperlein, E. M. and Short, R. W.},
  year = 1991,
  month = nov,
  journal = {Physics of Fluids B: Plasma Physics},
  volume = {3},
  number = {11},
  pages = {3092--3098},
  issn = {0899-8221},
  doi = {10.1063/1.859789},
  urldate = {2024-05-10}
}

@article{farmer2021,
  title = {Investigation of Heat Transport Using Directly Driven Gold Spheres},
  author = {Farmer, W. A. and Rosen, M. D. and Swadling, G. F. and Bruulsema, C. and Harris, C. D. and Rozmus, W. and Schneider, M. B. and Sherlock, M. W. and Edgell, D. H. and Katz, J. and Ross, J. S.},
  year = 2021,
  month = mar,
  journal = {Physics of Plasmas},
  volume = {28},
  number = {3},
  pages = {032707},
  issn = {1070-664X},
  doi = {10.1063/5.0040320},
  urldate = {2026-08-12}
}

@article{farmerHeatTransportModeling2018,
  title = {Heat Transport Modeling of the Dot Spectroscopy Platform on {{NIF}}},
  author = {Farmer, W A and Jones, O S and Barrios, M A and Strozzi, D J and Koning, J M and Kerbel, G D and Hinkel, D E and Moody, J D and Suter, L J and Liedahl, D A and Lemos, N and Eder, D C and Kauffman, R L and Landen, O L and Moore, A S and Schneider, M B},
  year = 2018,
  month = feb,
  journal = {Plasma Physics and Controlled Fusion},
  volume = {60},
  number = {4},
  pages = {044009},
  publisher = {IOP Publishing},
  issn = {0741-3335},
  doi = {10.1088/1361-6587/aaaefd}
}

@article{griff-mcmahon2024,
  title = {Measurements of Extended Magnetic Fields in Laser-Solid Interaction},
  author = {{Griff-McMahon}, J. and Malko, S. and {Valenzuela-Villaseca}, V. and Walsh, C. A. and Fiksel, G. and Rosenberg, M. J. and Schaeffer, D. B. and Fox, W.},
  year = 2024,
  month = sep,
  journal = {Physical Review Research},
  volume = {6},
  number = {3},
  pages = {033312},
  publisher = {American Physical Society},
  doi = {10.1103/PhysRevResearch.6.033312}
}

@inproceedings{hainesHeatFluxEffects1986,
  title = {Heat Flux Effects in {{Ohm}}'s Law},
  booktitle = {Plasma {{Physics}} and {{Controlled Fusion}}},
  author = {Haines, M G},
  year = 1986,
  volume = {28},
  pages = {1705--1716},
  doi = {10.1088/0741-3335/28/11/007}
}

@inproceedings{hohenbergerInertialConfinementFusion2012,
  title = {Inertial Confinement Fusion Implosions with Imposed Magnetic Field Compression Using the {{OMEGA Laser}}},
  booktitle = {Physics of {{Plasmas}}},
  author = {Hohenberger, M. and Chang, P. Y. and Fiksel, G. and Knauer, J. P. and Betti, R. and Marshall, F. J. and Meyerhofer, D. D. and S{\'e}guin, F. H. and Petrasso, R. D.},
  year = 2012,
  volume = {19},
  doi = {10.1063/1.3696032}
}

@inproceedings{joglekarMagneticReconnectionPlasma2014,
  title = {Magnetic Reconnection in Plasma under Inertial Confinement Fusion Conditions Driven by Heat Flux Effects in {{Ohm}}'s Law},
  booktitle = {Physical {{Review Letters}}},
  author = {Joglekar, a. S. and Thomas, a. G R and Fox, W. and Bhattacharjee, a.},
  year = 2014,
  volume = {112},
  pages = {1--5},
  doi = {10.1103/PhysRevLett.112.105004}
}

@article{jones2017,
  title = {Progress towards a More Predictive Model for Hohlraum Radiation Drive and Symmetry},
  author = {Jones, O. S. and Suter, L. J. and Scott, H. A. and Barrios, M. A. and Farmer, W. A. and Hansen, S. B. and Liedahl, D. A. and Mauche, C. W. and Moore, A. S. and Rosen, M. D. and Salmonson, J. D. and Strozzi, D. J. and Thomas, C. A. and Turnbull, D. P.},
  year = 2017,
  month = may,
  journal = {Physics of Plasmas},
  volume = {24},
  number = {5},
  pages = {056312},
  issn = {1070-664X},
  doi = {10.1063/1.4982693},
  urldate = {2026-08-12}
}

@article{kingham2002,
  title = {Nonlocal {{Magnetic-Field Generation}} in {{Plasmas}} without {{Density Gradients}}},
  author = {Kingham, R. J. and Bell, A. R.},
  year = 2002,
  month = jan,
  journal = {Physical Review Letters},
  volume = {88},
  number = {4},
  pages = {045004},
  publisher = {American Physical Society},
  doi = {10.1103/PhysRevLett.88.045004}
}

@article{liu2015,
  title = {Heat Flux Viscosity in Collisional Magnetized Plasmas},
  author = {Liu, C. and Fox, W. and Bhattacharjee, A.},
  year = 2015,
  month = may,
  journal = {Physics of Plasmas},
  volume = {22},
  number = {5},
  pages = {053302},
  issn = {1070-664X},
  doi = {10.1063/1.4918941},
  urldate = {2026-11-22}
}

@article{liu2017,
  title = {Momentum Transport and Nonlocality in Heat-Flux-Driven Magnetic Reconnection in High-Energy-Density Plasmas},
  author = {Liu, Chang and Fox, William and Bhattacharjee, Amitava and Thomas, Alexander G. R. and Joglekar, Archis S.},
  year = 2017,
  month = oct,
  journal = {Physical Review E},
  volume = {96},
  number = {4},
  pages = {043203},
  publisher = {American Physical Society},
  doi = {10.1103/PhysRevE.96.043203}
}

@article{luciani1983,
  title = {Nonlocal {{Heat Transport Due}} to {{Steep Temperature Gradients}}},
  author = {Luciani, J. F. and Mora, P. and Virmont, J.},
  year = 1983,
  month = oct,
  journal = {Physical Review Letters},
  volume = {51},
  number = {18},
  pages = {1664--1667},
  publisher = {American Physical Society},
  doi = {10.1103/PhysRevLett.51.1664}
}

@article{malone1975,
  title = {Indications of {{Strongly Flux-Limited Electron Thermal Conduction}} in {{Laser-Target Experiments}}},
  author = {Malone, R. C. and McCrory, R. L. and Morse, R. L.},
  year = 1975,
  month = mar,
  journal = {Physical Review Letters},
  volume = {34},
  number = {12},
  pages = {721--724},
  publisher = {American Physical Society},
  doi = {10.1103/PhysRevLett.34.721}
}

@article{manuelStudyHydrodynamicInstabilities2021,
  title = {On the Study of Hydrodynamic Instabilities in the Presence of Background Magnetic Fields in High-Energy-Density Plasmas},
  author = {Manuel, M. J.-E. and Khiar, B. and Rigon, G. and Albertazzi, B. and Klein, S. R. and Kroll, F. and Brack, F. -E. and Michel, T. and Mabey, P. and Pikuz, S. and Williams, J. C. and Koenig, M. and Casner, A. and Kuranz, C. C.},
  year = 2021,
  month = mar,
  journal = {Matter and Radiation at Extremes},
  volume = {6},
  number = {2},
  pages = {026904},
  issn = {2468-2047},
  doi = {10.1063/5.0025374},
  urldate = {2023-11-04}
}

@article{mikhailovskii1984,
  title = {Transport {{Equations}} of {{Plasma}} in a {{Curvilinear Magnetic Field}}},
  author = {Mikhailovskii, A. B. and Tsypin, V. S.},
  year = 1984,
  month = jan,
  journal = {Beitr\"age aus der Plasmaphysik},
  volume = {24},
  number = {4},
  pages = {335--354},
  publisher = {John Wiley \& Sons, Ltd},
  issn = {0005-8025},
  doi = {10.1002/ctpp.19840240404},
  urldate = {2026-11-21}
}

@article{ridgersMagneticCavitationReemergence2008,
  title = {Magnetic {{Cavitation}} and the {{Reemergence}} of {{Nonlocal Transport}} in {{Laser Plasmas}}},
  author = {Ridgers, C P and Kingham, R J and Thomas, A G R},
  year = 2008,
  month = feb,
  journal = {Physical Review Letters},
  volume = {100},
  number = {075003},
  doi = {10.1103/PhysRevLett.100.075003}
}

@inproceedings{ridgersMagneticFieldsNonLocal2008,
  title = {Magnetic {{Fields}} and {{Non-Local Transport}} in {{Laser Plasmas}}},
  booktitle = {Physics},
  author = {Ridgers, C.P.},
  year = 2008,
  number = {June}
}

@article{sadlerMagneticFieldGeneration2020,
  title = {Magnetic Field Generation from Composition Gradients in Inertial Confinement Fusion Fuel},
  author = {Sadler, James D. and Li, Hui and Flippo, Kirk A.},
  year = 2020,
  month = nov,
  journal = {Philosophical Transactions of the Royal Society A: Mathematical, Physical and Engineering Sciences},
  volume = {378},
  number = {2184},
  pages = {20200045},
  publisher = {Royal Society},
  doi = {10.1098/rsta.2020.0045},
  urldate = {2021-02-26}
}

@article{sadlerSymmetricSetTransport2021,
  title = {Symmetric {{Set}} of {{Transport Coefficients}} for {{Collisional Magnetized Plasma}}},
  author = {Sadler, James D. and Walsh, Christopher A. and Li, Hui},
  year = 2021,
  month = feb,
  journal = {Physical Review Letters},
  volume = {126},
  number = {7},
  pages = {075001},
  publisher = {American Physical Society},
  doi = {10.1103/PhysRevLett.126.075001}
}

@article{schurtz2000,
  title = {A Nonlocal Electron Conduction Model for Multidimensional Radiation Hydrodynamics Codes},
  author = {Schurtz, G. P. and Nicola{\"i}, {\relax Ph}. D. and Busquet, M.},
  year = 2000,
  month = oct,
  journal = {Physics of Plasmas},
  volume = {7},
  number = {10},
  pages = {4238--4249},
  issn = {1070-664X},
  doi = {10.1063/1.1289512},
  urldate = {2026-08-06}
}

@article{shaffer2023,
  title = {An Extended {{Vlasov}}--{{Fokker}}--{{Planck}} Approach for Kinetic Simulations of Laser Plasmas},
  author = {Shaffer, N. R. and Sherlock, M. and Maximov, A. V. and Goncharov, V. N.},
  year = 2023,
  month = apr,
  journal = {Physics of Plasmas},
  volume = {30},
  number = {4},
  pages = {043906},
  issn = {1070-664X},
  doi = {10.1063/5.0143248},
  urldate = {2026-09-25}
}

@article{sherlock2017,
  title = {A Comparison of Non-Local Electron Transport Models for Laser-Plasmas Relevant to Inertial Confinement Fusion},
  author = {Sherlock, M. and Brodrick, J. P. and Ridgers, C. P.},
  year = 2017,
  month = aug,
  journal = {Physics of Plasmas},
  volume = {24},
  number = {8},
  pages = {082706},
  issn = {1070-664X},
  doi = {10.1063/1.4986095},
  urldate = {2026-09-25}
}

@article{sherlockSuppressionBiermannBattery2020,
  title = {Suppression of the {{Biermann Battery}} and {{Stabilization}} of the {{Thermomagnetic Instability}} in {{Laser Fusion Conditions}}},
  author = {Sherlock, M. and Bissell, J. J.},
  year = 2020,
  month = feb,
  journal = {Physical Review Letters},
  volume = {124},
  number = {5},
  pages = {055001},
  publisher = {American Physical Society},
  doi = {10.1103/PhysRevLett.124.055001}
}

@article{shkarofsky1979,
  title = {Heat {{Conduction}} and {{Magnetic Field Induction}} in the {{Presence}} of {{Cold}}-and {{Hot-Electron Maxwellian Distributions}}},
  author = {Shkarofsky, I. P.},
  year = 1979,
  month = may,
  journal = {Physical Review Letters},
  volume = {42},
  number = {20},
  pages = {1342--1345},
  publisher = {American Physical Society},
  doi = {10.1103/PhysRevLett.42.1342}
}

@book{shkarofskyip1966,
  title = {The Particle Kinetics of Plasmas},
  author = {Shkarofsky, I. P. and Johnston, T. W. and Bachynski, M. P.},
  year = 1966,
  publisher = {Addison-Wesley},
  address = {Reading, MA}
}

@article{tzoufrasVlasovFokkerPlanck2011,
  title = {A {{Vlasov}}--{{Fokker}}--{{Planck}} Code for High Energy Density Physics},
  author = {Tzoufras, M. and Bell, A.R. and Norreys, P.A. and Tsung, F.S.},
  year = 2011,
  month = jul,
  journal = {Journal of Computational Physics},
  volume = {230},
  number = {17},
  pages = {6475--6494},
  issn = {0021-9991},
  doi = {10.1016/j.jcp.2011.04.034}
}

@misc{walsh_paper0,
  title = {Plasma Transport by Carrier Electrons: Classical Coefficients across Ionization and Magnetization},
  author = {Walsh, C. A.},
  year = 2026,
  eprint = {2610.30303},
  primaryclass = {physics.plasm-ph},
  archiveprefix = {arXiv}
}

@misc{walsh_paper1,
  title = {Plasma Transport by Carrier Electrons: Delocalization Length for Heat-Flux},
  author = {Walsh, C. A.},
  year = 2026,
  eprint = {2610.32085},
  primaryclass = {physics.plasm-ph},
  archiveprefix = {arXiv}
}

@misc{walsh_paper2,
      title={Plasma transport by carrier electrons: Nernst advection beyond the heat-flow analogy due to kinetic stress}, 
      author={C. A. Walsh},
      year={2026},
      eprint={2611.34766},
      archivePrefix={arXiv},
      primaryClass={physics.plasm-ph},
      url={https://arxiv.org/abs/2611.34766}, 
}

@article{walsh2024,
  title = {Resistive Diffusion in Magnetized {{ICF}} Implosions: {{Reduced}} Magnetic Stabilization of the {{Richtmyer}}--{{Meshkov}} Instability},
  author = {Walsh, C.A. and Strozzi, D.J. and Sio, H. and Pollock, B.B. and Appelbe, B.D. and Crilly, A.J. and O'Neill, S. and Weber, C. and Chittenden, J.P. and Moody, J.D.},
  year = 2024,
  month = jun,
  journal = {High Energy Density Physics},
  volume = {51},
  pages = {101103},
  issn = {1574-1818},
  doi = {10.1016/j.hedp.2024.101103}
}

@article{walshExtendedmagnetohydrodynamicsDensePlasmas2020,
  title = {Extended-Magnetohydrodynamics in under-Dense Plasmas},
  author = {Walsh, C. A. and Chittenden, J. P. and Hill, D. W. and Ridgers, C.},
  year = 2020,
  month = feb,
  journal = {Physics of Plasmas},
  volume = {27},
  number = {2},
  pages = {022103},
  publisher = {American Institute of Physics},
  issn = {1070-664X},
  doi = {10.1063/1.5124144},
  urldate = {2020-05-22}
}

@misc{walshKineticCorrectionsHeatflow2024,
  title = {Kinetic {{Corrections}} to {{Heat}}-Flow and {{Nernst}} Advection for {{Laser Heated Plasmas}}},
  author = {Walsh, Chris and Sherlock, Mark},
  year = 2024
}

\end{document}